\documentclass[aps,pra,twocolumn,amsmath,amssymb,superscriptaddress,longbibliography]{revtex4-1}

\usepackage{mathrsfs}
\usepackage{graphicx}
\usepackage{amsmath}
\usepackage{amssymb}
\usepackage{amsfonts}
\usepackage{color}
\usepackage{CJK}
\usepackage{appendix}

\usepackage{multirow}
\usepackage{booktabs}
\usepackage{bigstrut}

\usepackage{lipsum}

\usepackage[unicode=true,colorlinks=true,citecolor=blue,urlcolor=blue]{hyperref}

\def\hatd#1{\hat{#1}^\dagger}

\def\ket#1{\left|{#1}\right\rangle}
\def\braket#1#2{\left\langle{{#1}}\mathrel{\left|{\vphantom{{#1}{#2}}}\right.\kern-\nulldelimiterspace}{{#2}}\right\rangle}

\begin{document}

\title{Floquet Engineering of Hilbert Space Fragmentation in Stark Lattices}

\author{Li Zhang}
\affiliation{Institute of Quantum Precision Measurement, State Key Laboratory of Radio Frequency Heterogeneous Integration, Shenzhen University, Shenzhen 518060, China}
\affiliation{College of Physics and Optoelectronic Engineering, Shenzhen University, Shenzhen 518060, China}

\author{Yongguan Ke}
\affiliation{Laboratory of Quantum Engineering and Quantum Metrology, School of Physics and Astronomy, Sun Yat-Sen University (Zhuhai Campus), Zhuhai 519082, China}
\affiliation{Institute of Quantum Precision Measurement, State Key Laboratory of Radio Frequency Heterogeneous Integration, Shenzhen University, Shenzhen 518060, China}
\affiliation{College of Physics and Optoelectronic Engineering, Shenzhen University, Shenzhen 518060, China}

\author{Ling Lin}
\affiliation{Institute of Quantum Precision Measurement, State Key Laboratory of Radio Frequency Heterogeneous Integration, Shenzhen University, Shenzhen 518060, China}
\affiliation{College of Physics and Optoelectronic Engineering, Shenzhen University, Shenzhen 518060, China}

\author{Chaohong Lee}
\altaffiliation{Email: chleecn@szu.edu.cn}
\affiliation{Institute of Quantum Precision Measurement, State Key Laboratory of Radio Frequency Heterogeneous Integration, Shenzhen University, Shenzhen 518060, China}
\affiliation{College of Physics and Optoelectronic Engineering, Shenzhen University, Shenzhen 518060, China}
\affiliation{Quantum Science Center of Guangdong-Hongkong-Macao Greater Bay Area (Guangdong), Shenzhen 518045, China}

\begin{abstract}
The concept of Hilbert space fragmentation (HSF) has recently been put forward as a routine to break quantum ergodicity.
While HSF widely exists in dynamical constraint models, it is still challenging to tune HSF.
Here, we propose a scheme to tune HSF in a one-dimensional tilted lattice of interacting spinless fermions with periodically driven tunneling.
The dynamics is governed by effective Hamiltonians with kinetic constraints, which appear as density-dependent tunneling in the weak-tunneling perturbation expansion.
The kinetic constraint can be tuned via changing the driving frequency, and three different kinds of strong HSF can be engineered.
In general, the system is strongly constrained and exhibits a strong HSF.
Two partial resonance frequencies are \emph{analytically} given by a time-dependent perturbation theory for Floquet systems, at which some kinetic constraints are released and the system exhibits another two different strong HSF.
We demonstrate the perturbation analysis with exact numerical simulation of the entanglement entropy, the density correlation functions and the saturated local density profiles.
Our result provides a promising way to control HSF through Floquet engineering.
\end{abstract}

\date{\today}
\maketitle

\section{Introduction\label{Sec1}}
Recently, there has been much interests in exploring whether and how an isolated quantum many-body system can reach thermal equilibrium under unitary dynamics.
While ergodic systems can reach thermal equilibrium rapidly via eigenstate thermalization hypothesis~\cite{Deutsch1991,Srednicki1994,Srednicki1999,Rigol2008,Abanin2016}, the failure of thermalization due to ergodicity breaking has been predicted in several systems~\cite{Nandkishore2015,Abanin2019,Moudgalya2022}.
Two well-known examples for ergodicity breakdown are quantum integrable systems and many-body localized systems, both of which possess an extensive number of conserved quantities~\cite{Kinoshita2006,Rigol2007,Polkovnikov2011,Serbyn2013,Huse2014,Nandkishore2015,Abanin2019}.
Moreover, exotic violation of ergodicity has been found in systems subjected to strong tilting potentials~\cite{Schulz2019,Nieuwenburg2019,Taylor2020,Yao2020,Chanda2020,Zhang2021,Yao2021,Wang2021,Guo2021,Morong2021}, which support quantum many-body scars~\cite{Turner2018,Turner2018_2,Ho2019,Serbyn2021,Moudgalya2022,Chandran2023} and fractured Hilbert space~\cite{Pai2019,Sala2020,Khemani2020,Moudgalya2021,Moudgalya2022}.

The Hilbert space fragmentation (HSF) describes the phenomenon that the Hilbert space of a system is exponentially split into many dynamically disjoint invariant subspaces (referred to as Krylov subspaces), which can not be captured by the conventional symmetry~\cite{Pai2019,Sala2020,Khemani2020,Moudgalya2021,Moudgalya2022}.
Thus, large parts of the Hilbert space are inaccessible to certain initial states and the ergodicity breaks down regarding the full system.
The HSF arises due to dynamical constraints, such as fractonlike constraints of conservation of U(1) charge and its associated dipole~\cite{Pai2019,Sala2020,Khemani2020,Moudgalya2021} and local tunneling constraints from strong interaction or strong tilting potentials~\cite{Tomasi2019,Yang2020,Frey2022,Ghosh2023,Kohlert2021,Kohlert2023}.
Depending on whether the largest Krylov subspace can span almost the entire symmetry space or not in the thermodynamic limit, the HSF is categorized as weak and strong respectively~\cite{Sala2020,Moudgalya2022}.
Although the HSF can be different when changing the filling number~\cite{Morningstar2020,Pozderac2023}, to our best knowledge, there is still no way to engineer the HSF by tuning the physical parameters.

Floquet engineering, the coherent control of a quantum system via periodic driving, is a powerful tool for engineering synthetic Hamiltonians with novel properties~\cite{Eckardt2017}.
It has been used to realize topological Bloch bands, dynamical localization and synthetic gauge field, etc~\cite{Eckardt2017}.
Specifically, density-dependent tunneling has been engineered by suitable driving schemes~\cite{Eckardt2017,Greschner2014,Meinert2016,Zhao2019}.
Recently, it have been proposed to generate quantum many-body scars and HSF via tuning local tunneling constraints to some specific parameters~\cite{Zhao2020,Ghosh2023}.
However, the way to adjust the HSF via Floquet engineering is still lacking.

In this article, we provide a way to tune the HSF in a periodically driven Stark chain of interacting spinless fermions.
The fermions with nearest neighboring interactions are subjected to a tilted field, and they tunnel between nearest neighboring sites with periodically varying tunneling strength.
When the tunneling strength is much smaller than the tilting strength, the interaction strength and the driving frequency, the system at stroboscopic times can be viewed as transitions between multiple energy levels by absorbing or emitting ``photons'' from the driving field.
The energy levels are formed by the eigenergies of the tilting and interacting Hamiltonian.
The transitions between different energy levels happen through the action of the hopping Hamiltonian, with frequencies and amplitudes given by the Fourier expansion of the tunneling strength.
Thus, there appears a resonant transition when the Fourier frequencies match the energy difference before and after the hopping, which depends on the particle numbers near the hopping sites.
The non-resonant transitions can be neglected in the presence of resonant ones.
Inspired by this picture, we can tune the relation between the driving frequency and the interaction and tilting strength to design different Hamiltonians with different kinetic constraints, in which some particular hoppings are non-resonant thus forbidden.
As a concrete example, we set the interaction and tilting strength being equal, at which an intrinsic resonance is always present, and tune the driving frequency to obtain other resonant transitions.
In the absence of the driving, the hopping is constrained to conserve the total energy of interaction and tilting and leads to strong HSF.
In the presence of the periodic driving, we engineer two extra resonant hoppings at two particular resonant frequencies respectively, which are analytically given by a perturbation method for Floquet system.
The released kinetic constraint changes the structure of the Hilbert space and leads to two different kinds of HSF.
At non-resonant frequencies, the driving does not assist the hopping, and the HSF is the same as that in the absence of the periodic driving.
We demonstrate the perturbation analysis through exact numerical calculation of the dynamics of entanglement entropy, the density autocorrelation function and the saturated density profiles.
Our result provides a route to controlled switching between different HSF through tuning the driving frequency.

The rest of the paper is organized as follows.
In Sec.~\ref{Sec2}, we introduce the model for our physical system and present a time-dependent perturbation theory for Floquet system to derive the effective Hamiltonians.
In Sec.~\ref{Sec3}, we present how to engineer the HSF by tuning the driving frequency.
At last, in Sec.~\ref{Sec4}, we give a brief summary of our results.

\section{Model and Method}\label{Sec2}

\subsection{Model}
\begin{figure}[t]
\includegraphics[width=\columnwidth]{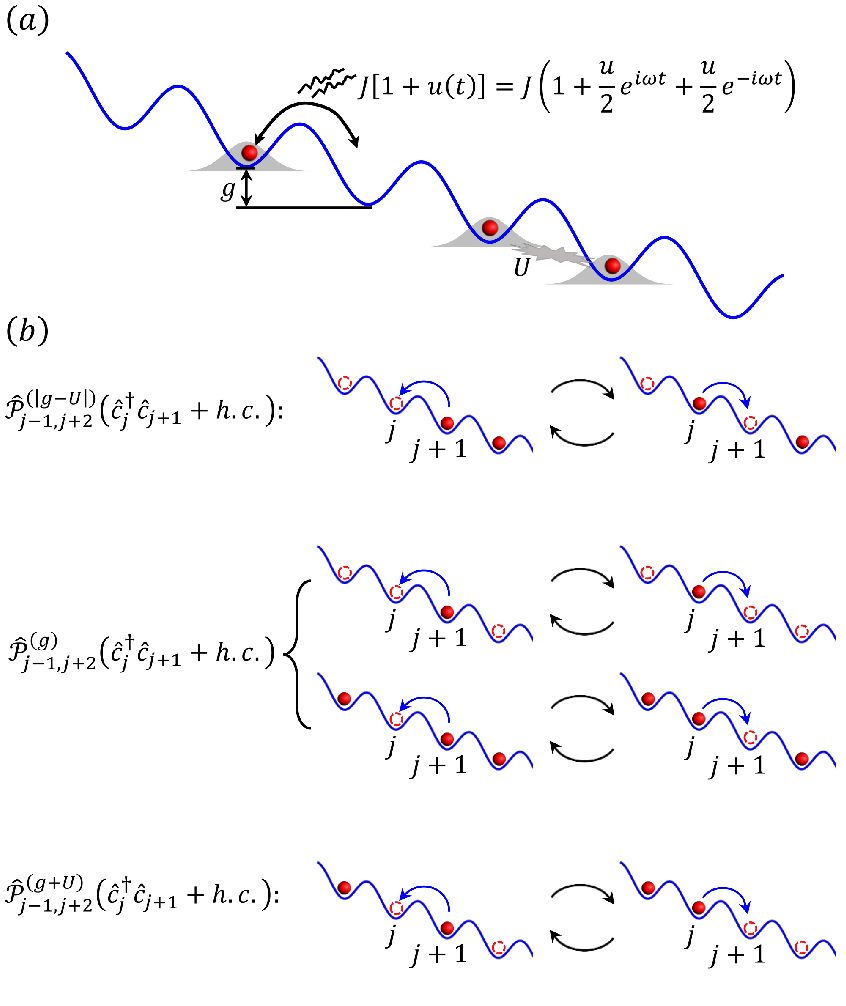}
  \caption{\label{Fig_potential}
  (a) Schematic diagram of interacting spinless fermions in a periodically driven tilted lattice.
  The tilting and nearest neighboring interaction strength are $g$ and $U$ respectively.
  The fermions tunnel between nearest neighboring sites with periodic varying strength $J[1+u(t)]$, where $u(t)=u\cos \omega t$.
  (b) Illustration of the three tunneling processes, which are characterized by the projected tunnelings $\hat{\mathcal P}^{(|\Delta_i|)}(\hatd c_j\hat c_{j+1}+h.c.)$ ($i=1,2,3$) with $|\Delta_1|=|g-U|$, $|\Delta_2|=g$ and $|\Delta_3|=g+U$.
  The red balls denote the fermions and the red dashed circles denotes the empty sites.
  }
\end{figure}

We consider an ensemble of interacting spinless fermions in a one-dimensional Stark lattice under periodic driving, see Fig.~\ref{Fig_potential}.
It is described by the Hamiltonian $\hat H(t)=\hat H_0+\hat H_\mathrm{d}(t)$
with the time-independent part
\begin{equation}\label{Eq.H0}
\hat H_0=\hat H_J+U\sum_{j=0}^{L-2}\hat n_j\hat n_{j+1}-g\sum_{j=0}^{L-1}j\hat n_j,
\end{equation}
where the hopping term $\hat H_J=J\sum_{j=0}^{L-2}(\hatd c_j\hat c_{j+1}+\hatd c_{j+1}\hat c_j)$,
and the time periodic part
\begin{equation}
\hat H_\mathrm{d}(t)=u(t)\hat H_J.
\end{equation}
Here, $\hatd c_j (\hat c_j)$ creates (annihilates) a fermion at site $j$, $\hat n_j=\hatd c_j\hat c_j$ is the particle number operator, and $u(t)=u\cos(\omega t)$ is periodic in time with frequency $\omega$ and amplitude $u$.
The parameters $J$, $U$ and $g$ are the nearest-neighbor tunneling strength, the nearest-neighbor interaction strength and the tilting field strength, respectively.
$L$ is the total number of lattice sites.
In Appendix~\ref{AppA}, we provide a discussion of the realization of our model in Rydberg atom platform.
Our model may also be simulated by insulating two-component bosonic atoms trapped in a one-dimensional tilted optical lattice~\cite{Liu2020}.
In the following study, we consider open boundary condition and focus on the half-filling sector with particle number $N=L/2$.
We set $\hbar=1$ and the energy unit as $J=1$.

\subsection{Time dependent perturbation theory in Floquet systems}

Generally, the dynamics of a periodically driven system $\hat H(t)$ after a period is governed by its Floquet Hamiltonian
\begin{equation}
\hat H_F=\frac{i}{T}\ln\hat F.
\end{equation}
Here the Floquet operator $\hat F$ is the unitary evolution operator over a period:
\begin{equation}
\hat F=\mathcal{T}\mathrm{exp}\left[-i\int_0^{T}\hat H(t)dt\right],
\end{equation}
with $\mathcal{T}$ denoting time ordering.
In general cases, $\hat H_F$ is untractable, as $\hat H(t)$ at different times does not commute with each other.
Here, we apply the time-dependent perturbation theory (TDPT) in Floquet systems to obtain the Floquet Hamiltonian for small driving amplitudes~\cite{Soori2010,Sen2021}.

In our model, when $J, uJ\ll g,U,\omega/2\pi$, and $g$ and $U$ are comparable, we can treat $[1+u(t)]\hat H_J$ as a perturbation to $U\sum_{j=0}^{L-2}\hat n_j\hat n_{j+1}-g\sum_{j=0}^{L-1}j\hat n_j$.
The first-order effective Floquet Hamiltonian is derived in Appendix.
The first-order perturbation process can be understood from the following physical picture.
Since $U$ and $g$ are much larger than the tunneling strength, tunneling is suppressed due to large energy barrier $\Delta$.
On the other hand, the tunneling Hamiltonian can be written as $[1+u(t)]\hat H_J=(1+0.5ue^{i\omega t}+0.5ue^{-i\omega t})\hat H_J$, in which the periodic part can be viewed as ``photons'' assisted tunneling with ``photon'' energy $\omega$.
So, the tunneling can be stimulated if $\Delta=0$ or $|\Delta|=\omega$, i.e. under the condition of resonance between the interaction and tilting potential or resonance between the system and the driving.
In our model, there are three tunneling channels characterized by energy barriers $|\Delta_i|=|g-U|$, $g$ and $g+U$ ($i=1,2,3$), which depend on the particle density near the hopping sites, see Fig.~\ref{Fig_potential}(b).
They can be depicted by projected tunnelings $\hat{\mathcal P}^{(|\Delta_i|)}_{j-1,j+2}(\hatd c_{j+1}\hat c_j+h.c.)$ with projectors
\begin{eqnarray}
\hat{\mathcal P}_{j-1,j+2}^{(|g-U|)}&=&\hat n_{j+2}(1-\hat n_{j-1}),\nonumber\\
\hat{\mathcal P}_{j-1,j+2}^{(g)}&=&1-(\hat n_{j-1}-\hat n_{j+2})^2, \nonumber\\
\hat{\mathcal P}_{j-1,j+2}^{(g+U)}&=&\hat n_{j-1}\left(1-\hat n_{j+2}\right)
\end{eqnarray}

We can tune the relation between the interaction, tilting strength and the driving frequency to make particular tunneling process resonant.
To reduce flexibility, we fix $U=g$.
The first-order Floquet Hamiltonian in this case reads
\begin{widetext}
\begin{eqnarray}\label{Eq.FirstOrderHF}
\hat H_F^{(1)}&=&\frac{i}{T}\ln e^{-i\hat E T}+J\sum_j\hat{\mathcal P}^{(0)}_{j-1,j+2}\hatd c_{j+1}\hat c_j+h.c.\nonumber\\
&+&\left\{\frac{iJ}{T}\left(e^{-igT}-1\right)\left[\frac{1}{g}+\frac{u\left(1-\delta_{g,\omega}\right)}{2(g-\omega)} +\frac{u}{2\left(g+\omega\right)}\right]+\frac{uJ}{2}\delta_{g,\omega}\right\}\sum_j\hat P^{(g)}_{j-1,j+2}\hatd c_{j+1}\hat c_j+h.c.\nonumber\\
&+&\left\{\frac{iJ}{T}\left(e^{-i2gT}-1\right)\left[\frac{1}{2g}+\frac{u\left(1-\delta_{2g,\omega}\right)}{2(2g-\omega)} +\frac{u}{2\left(2g+\omega\right)}\right]+\frac{uJ}{2}\delta_{2g,\omega}\right\}\sum_j\hat P^{(2g)}_{j-1,j+2}\hatd c_{j+1}\hat c_j+h.c.,
\end{eqnarray}
\end{widetext}
where $\hat{\mathcal P}^{(0)}_{j-1,j+2}=\hat{\mathcal P}^{(|g-U|)}_{j-1,j+2}$ and $\hat{\mathcal P}^{(2g)}_{j-1,j+2}=\hat{\mathcal P}^{(g+U)}_{j-1,j+2}$.

\begin{widetext}
\begin{eqnarray}
\hat H_F^{(1)}&=&\frac{i}{T}\ln e^{-i\hat ET}\nonumber\\
&+&J\left\{\delta_{g,U}+\frac{u}{2}\delta_{|g-U|,\omega}+ (1-\delta_{g,U})\frac{i\left[e^{-i(g-U)T}-1\right]}{T}\left[\frac{1}{g-U}+\frac{u(1-\delta_{|g-U|,\omega})(g-U)}{(g-U+\omega)(g-U-\omega)}\right]\right\}\sum_j\hat{\mathcal P}_{j-1,j+2}^{(g-U)}\hatd c_{j+1}\hat c_j+h.c. \nonumber\\
&+&J\left\{\frac{u}{2}\delta_{g,\omega}+ \frac{i\left(e^{-igT}-1\right)}{T}\left[\frac{1}{g}+\left(1-\delta_{g,\omega}\right)\frac{ug}{(g+\omega)(g-\omega)}\right]\right\}\sum_j\hat{\mathcal P}_{j-1,j+2}^{(g)}\hatd c_{j+1}\hat c_j+h.c.\nonumber\\
&+&J\left\{\frac{u}{2}\delta_{g+U,\omega}+\frac{i\left[e^{-i(g+U)T}-1\right]}{T}\left[\frac{1}{g+U}+(1-\delta_{g+U,\omega})\frac{u(g+U)}{(g+U+\omega)(g+U-\omega)}\right]\right\}\hat{\mathcal P}_{j-1,j+2}^{(g+U)}\hatd c_{j+1}\hat c_j+h.c.
\end{eqnarray}
\end{widetext}

Consider a Hamiltonian $\hat H(t)=\hat H_s+\lambda \hat H_V(t)$, where $\hat H_s$ is time independent and solvable with eigenequation $\hat H_s|n\rangle=E_n|n\rangle$, $\hat H_V(t)$ is time periodic with frequency $\omega$ and period $T=2\pi/\omega$, and $\lambda \ll 1$.
We now find solution to the time-dependent Schr\"{o}dinger equation (TDSE) $i\partial_t\ket{\Psi(t)}=\hat H(t)\ket{\Psi(t)}$ with nonzero $\lambda$, where $\ket{\Psi(t)}$ should satisfy
\begin{equation}\label{Eq.EvolutionEq}
|\Psi(T)\rangle=\hat F|\Psi(0)\rangle,
\end{equation}
according to the definition of the Floquet operator $\hat F$.
When $\lambda=0$, $|n(t)\rangle\equiv e^{-iE_nt}|n\rangle$ satisfies the TDSE $i\partial_t|n(t)\rangle=\hat H_s|n(t)\rangle$.
When $0<\lambda\ll1$, we expand $|\Psi(t)\rangle$ in the basis of $|\{n(t)\rangle\}$
\begin{equation}
|\Psi(t)\rangle=\sum_nc_n(t)e^{-iE_nt}|n\rangle.
\end{equation}
Substituting it to the TDSE yields
\begin{equation}\label{Eq.CoeffRelation}
c_n(t)=c_n(0)-i\lambda\sum_l\int_0^t c_l(t')e^{-i\Delta_{l,n}t'}V_{n,l}(t')dt',
\end{equation}
where $\Delta_{l,n}=E_l-E_n$ is the energy difference between the states $|l\rangle$ and $|n\rangle$, and $V_{n,l}(t)=\langle n|\hat H_V(t)|l\rangle$ is the matrix elements of $\hat H_V(t)$ in the basis $\{|n\rangle\}$.
When $\lambda \ll 1$, we expand $c_n(t)$ in powers of $\lambda$ as
\begin{equation}
c_n(t)=c_n^{(0)}+\lambda c_n^{(1)}(t)+\lambda^2 c_n^{(2)}(t)+\mathcal O(\lambda^3).
\end{equation}
Through substituting it into Eq.\eqref{Eq.CoeffRelation} and comparing the coefficients before the $k$th order of $\lambda$ of both sides of the equation, we obtain the series of coefficients
\begin{widetext}
\begin{eqnarray}
&&c_n^{(1)}(t)=c_n^{(1)}(0)-i\sum_l\int_0^tc_l^{(0)}e^{-i\Delta_{l,n}t_1}V_{n,l}(t_1)dt_1, \nonumber\\
&&c_n^{(2)}(t)=c_n^{(2)}(0)-i\sum_l\int_0^tc_l^{(1)}(t_1)e^{-i\Delta_{l,n}t_1}V_{n,l}(t_1)dt_1\nonumber\\
&&\qquad\quad=c_n^{(2)}(0)-i\sum_l\int_0^tc_l^{(1)}(0)e^{-i\Delta_{l,n}t_1}V_{n,l}(t_1)dt_1-\sum_{l,m_1}\int_0^t \left[\int_0^{t_1}c_l^{(0)}e^{-i\Delta_{l,m_1}t_2}V_{m_1,l}(t_2)dt_2\right]e^{-i\Delta_{m_1,n}t_1}V_{n,m_1}(t_1)dt_1,\nonumber\\
&&\cdots.
\end{eqnarray}
Thus,
\begin{eqnarray}\label{Eq.Coeff}
c_n(t)&=&c_n^{(0)}+\lambda c_n^{(1)}(0)+\lambda^2c_n^{(2)}(0)-i\lambda\sum_l\int_0^t\left[ c_l^{(0)}+\lambda c_l^{(1)}(0)\right]e^{-i\Delta_{l,n}t_1}V_{n,l}(t_1)dt_1\nonumber\\
&&\qquad-\lambda^2\sum_{l,m_1}\int_0^t\left[\int_0^{t_1}c_l^{(0)}e^{-i\Delta_{l,m_1}t_2}V_{m_1,l}(t_2)dt_2\right] e^{-i\Delta_{m_1,n}t_1}V_{n,m_1}(t_1)dt_1+\mathcal O (\lambda^3)\nonumber\\
&=&c_n(0)-i\lambda\sum_lc_l(0)\int_0^te^{-i\Delta_{l,n}t_1}V_{n,l}(t_1)dt_1\nonumber\\
&&\qquad\quad-\lambda^2 \sum_{l,m_1}c_l(0)\int_0^t\left[\int_0^{t_1}e^{-i\Delta_{l,m_1}t_2}V_{m_1,l}(t_2)dt_2\right]e^{-i\Delta_{m_1,n}t_1}V_{n,m_1}(t_1)dt_1+\mathcal O (\lambda^3).
\end{eqnarray}
Labeling the column vector $[c_1(t);c_2(t);\cdots;c_{\mathcal D}(t)]$ as $|c(t)\rangle$ with $\mathcal D$ being the total Hilbert space dimension, Eq.\eqref{Eq.Coeff} at $t=T$ can be written in the matrix form
\begin{equation}
|c(T)\rangle=[\hat I+\lambda\hat F_1+\lambda^2\hat F_2+\mathcal O(\lambda^3)]|c(0)\rangle,
\end{equation}
where $\hat I$ is the identity matrix and
\begin{eqnarray}\label{Eq.F1F2}
\hat F_1&=&-i\sum_{n,l}\int_0^Te^{-i\Delta_{l,n}t_1}V_{n,l}(t_1)dt_1|n\rangle\langle l|,\nonumber\\
\hat F_2&=&-\sum_{n,l}\sum_{m_1}\int_0^Te^{-i\Delta_{m_1,n}t_1}V_{n,m_1}(t_1)\left[\int_0^{t_1}e^{-i\Delta_{l,m_1}t_2}V_{m_1,l}(t_2)dt_2\right]dt_1|n\rangle\langle l|.
\end{eqnarray}
\end{widetext}
Writing Eq.\eqref{Eq.EvolutionEq} in the matrix form
\begin{equation}
e^{-i\hat ET}|c(T)\rangle=\hat F|c(0)\rangle,
\end{equation}
where $e^{-i\hat ET}=\mathrm{diag}([e^{-iE_1T} \ e^{-iE_2T}\ \cdots \ e^{-iE_{\mathcal D}T}])$, one can directly write down the Floquet operator
\begin{equation}
\hat F=e^{-i\hat ET}\left[\hat I+\lambda\hat F_1+\lambda^2\hat F_2+\mathcal O(\lambda^3)\right].
\end{equation}

The $k$th order Floquet operator $\lambda^k\hat F_k$ can be viewed as the transition between levels $l$ and $n$ intermediated by $k$ virtual transitions assisted by the driving.
For each transition between levels $m_i$ and $m_{i+1}$, expanding the periodic matrix element in the Fourier series as $V_{m_{i+1},m_i}(t)=\sum_{q_i}f_{q_i}(m_i,m_{i+1})e^{iq_i\omega t}$ ($q_i$ being integer numbers), the driving provides multiple frequencies $q_i\omega$ if $f_{q_i}$ is nonzero.
If there exists a $\tilde q_i$ making $\tilde q_i \omega=\Delta_{m_i,m_{i+1}}$, single-photon resonance occurs between $m_i$ and $m_{i+1}$.
For a $p$ order transition between levels $m_i$ and $m_{i+p}$, if all intermediate single-photon processes are non-resonant but there exists a set of $\{\tilde q_i, \tilde q_{i+1}, \cdots, \tilde q_{i+p-1}\}$ making $\Delta_{m_i,m_{i+p}}=(\tilde q_i+\tilde q_{i+1}+\cdots+\tilde q_{i+p-1})\omega$, $p$-photon resonance between $m_i$ and $m_{i+p}$ occurs.
If $\hat F_k$ contains $q_r (0\leq q_r\leq k)$ resonance processes, no matter single-photon or multi-photon resonance, the integrals over time lead to a factor $T^{q_r}$.
Thus, in a tight manner, if $\lambda T\ll1$, one can expand the Floquet Hamiltonian in the Taylor series as
\begin{eqnarray}\label{Eq.EffHam}
&&\hat H_F=\frac{i}{T}\ln \left\{e^{-i\hat ET}\left[\hat I+\lambda\hat F_1+\lambda^2\hat F_2+\mathcal O(\lambda^3)\right]\right\} \nonumber\\
&&=\hat H_{\omega}+\frac{i}{T}\left[\lambda\hat F_1+\lambda^2\hat F_2-\frac{\lambda^2}{2}\hat F_1^2+\mathcal O(\lambda^3 T^3)\right],
\end{eqnarray}
where $\hat H_{\omega}=\frac{i}{T}\ln e^{-i\hat ET}=\mathrm{diag}\{\mathrm{mod}([E_1\ E_2 \ \cdots E_{\mathcal D}],\omega)\}$ is the effective on-site potential by folding the unperturbed energies into one Floquet Brilloun zone with width $\omega$.

For our model, we treat $\lambda\hat H_V(t)=[1+u(t)]\hat H_J$ as perturbation to $\hat H_s=U\sum_{j=0}^{L-2}\hat n_j\hat n_{j+1}-g\sum_{j=0}^{L-1}j\hat n_j$, assuming $J,u\ll g,U,\omega/2\pi$, and $g$ and $U$ are comparable.
The eigenstates of $\hat H_s$ are the Fock states $|\vec n\rangle=|n_0 n_1\cdots n_{L-2}n_{L-1}\rangle$ with eigenergy $E_{\vec n}=U\sum_{j=0}^{L-2} n_j n_{j+1}-g\sum_{j=0}^{L-1}j n_j$, where $n_j=0,1$ denotes the particle number on site $j$.
We consider perturbation to first order.
The nonzero tunneling matrix element between $|\vec n\rangle$ and $|\vec l\rangle$ can be expanded in the Fourier series as $V_{\vec n,\vec l}(t)=J(1+0.5ue^{i\omega t}+0.5ue^{-i\omega t})$.
Thus, resonance occurs if $\Delta_{\vec l,\vec n}=0$ and $\pm \omega$.
In our model, there are three different pairs of conjugate tunneling processes: a particle tunnels to the right/left site with the total interaction strength unchanged, increased/decreased and decreased/increased by $U$, which cause energy difference $|\Delta|=g,|g-U|$ and $|g+U|$ respectively.
Thus, we can tune the relation between the interaction, tilting strength and the driving frequency to make particular tunneling process resonant.

To reduce flexibility, we fix $U=g$, which ensures the intrinsic resonant tunneling without driving.
The first order Floquet Hamiltonian in this case reads
\begin{widetext}
\begin{eqnarray}\label{Eq.FirstOrderHF}
\hat H_F^{(1)}&=&\frac{i}{T}\ln e^{-i\hat E T}+J\sum_j\hat{\mathcal P}^{(0)}_{j-1,j+2}\hatd c_{j+1}\hat c_j+h.c.\nonumber\\
&+&\left\{\frac{iJ}{T}\left(e^{-igT}-1\right)\left[\frac{1}{g}+\frac{u\left(1-\delta_{g,\omega}\right)}{2(g-\omega)} +\frac{u}{2\left(g+\omega\right)}\right]+\frac{uJ}{2}\delta_{g,\omega}\right\}\sum_j\hat P^{(g)}_{j-1,j+2}\hatd c_{j+1}\hat c_j+h.c.\nonumber\\
&+&\left\{\frac{iJ}{T}\left(e^{-i2gT}-1\right)\left[\frac{1}{2g}+\frac{u\left(1-\delta_{2g,\omega}\right)}{2(2g-\omega)} +\frac{u}{2\left(2g+\omega\right)}\right]+\frac{uJ}{2}\delta_{2g,\omega}\right\}\sum_j\hat P^{(2g)}_{j-1,j+2}\hatd c_{j+1}\hat c_j+h.c.,
\end{eqnarray}
\end{widetext}
where
\begin{eqnarray}
\hat{\mathcal P}_{j-1,j+2}^{(0)}&=&\hat n_{j+2}(1-\hat n_{j-1}),\nonumber\\
\hat{\mathcal P}_{j-1,j+2}^{(g)}&=&1-(\hat n_{j-1}-\hat n_{j+2})^2, \nonumber\\
\hat{\mathcal P}_{j-1,j+2}^{(2g)}&=&\hat n_{j-1}\left(1-\hat n_{j+2}\right)
\end{eqnarray}
accounts for the tunneling processes with energy difference being 0, $g$ and $2g$ respectively.
Apart from the intrinsic resonance proportional to $\hat{ \mathcal P}^{(0)}_{j-1,j+2}$, there can be two partial resonances between the driving and the system at frequencies $\omega_1=g$ and $\omega_2=2g$ respectively, where the tunneling processes proportional to $\hat {\mathcal P}_{j-1,j+2}^{(g)}$ and $\hat{\mathcal P}_{j-1,j+2}^{(2g)}$ is resonant correspondingly.
At other frequencies, the system does not resonate with the driving.
Particularly, at $\omega=g/q$ (with $q$ being integers larger than 1), the driving does not assist the tunneling and the two tunneling processes vanish.
In the following study, we show that the HSF can be tuned by changing the driving frequency to and away from the resonance frequencies.

\section{Floquet Engineering of Hilbert space fragmentation}\label{Sec3}
In this section, we explore the tuning of HSF by controlling the driving frequency within the half-filling sector.
We explore the splitting of the Hilbert space based on $\hat H_F^{(1)}$, and supplement the results by the exact dynamical evolution governed by the evolution operator $\hat U(t)=\exp[-i\hat H_0t]$ in the absence of driving, and the Floquet operator $\hat F=\mathcal T\int_0^T\mathrm{exp}[-i\hat H(t)t]dt$ in the presence of driving.
The system size is set as $L=16$ unless otherwise specified.

Some of the key diagnostics used to identify the HSF include the saturated values of bipartite entanglement entropy (EE), infinite-temperature autocorrelation functions and local observables.

The bipartite von Neumann EE between a subsystem $A$ and the rest of the system $B$ is defined as
\begin{equation}
S=-\mathrm{Tr}\left(\hat \rho_A\ln \hat \rho_A\right),
\end{equation}
where $\hat \rho_A=\mathrm{Tr}_B(|\psi\rangle\langle \psi|)$ is the reduced density matrix of the subsystem $A$.
When evolving from a low-entangled initial state $|\psi_0\rangle$ in a finite system, the EE will saturate to a value $S_{\infty}$.
If the Krylov subspace to which the initial state belongs is ergodic, $S_{\infty}$ is consistent with the Page value of that subspace~\cite{Page1993,Yang2020,Ghosh2023}.
For a system without HSF, the Krylov subspace is trivially the symmetry sector.
The Page value $S_p$ of the half-filling sector can be calculated as
\begin{equation}
S_p=-\sum_{n_A=0}^{N}\frac{d_A(n_A)d_B(N-n_A)}{\mathcal D}\ln \frac{d_B(N-n_A)}{\mathcal D}-\frac{1}{2},
\end{equation}
where $\mathcal D$ is the dimension of the half-filling sector and $d_A(n_A)[d_B(N-n_A)]$ is the dimension of the subsystem $A[B]$ with particle number $n_A[N-n_A]$~\cite{Ghosh2023}.
For a system with HSF, the Page value $S_p[\mathcal K]$ in a Krylov subspace $\mathcal K$ can be calculated via averaging the von Neumann EE of the random canonical states in $\mathcal K$
\begin{equation}\label{Eq.CanoRandomSts}
|\phi\rangle=\sum_{n=1}^{\mathcal D_{\mathcal K}}\frac{z_{n}}{\sqrt{\mathcal N}}|n\rangle,
\end{equation}
where $\{|n\rangle\}$ is the set of basis in $\mathcal K$, $\mathcal D_{\mathcal K}$ is the dimension of $\mathcal K$, $z_n$ are normally distributed real random numbers with mean 0 and variance 1 and $\sqrt{\mathcal N}$ is the normalized coefficient~\cite{Ghosh2023}.
In the following study, we consider the half-chain von Neumann EE with $A$ being the left half chain of the system, and calculate $S_p[\mathcal K]$ in a Krylov subspace $\mathcal K$ by averaging the EE over $\mathcal D_{\mathcal K}$ random canonical states.

The infinite-temperature autocorrelation function of an operator $\hat O$ is defined as $C(t)=\langle\psi_{\mathrm{inf}}|\hat O(t)\hat O(0)|\psi_{\mathrm{inf}}\rangle$, where $|\psi_{\mathrm{inf}}\rangle$ is a random infinite-temperature state distributing evenly on the whole Hilbert space.
It captures the spreading of operator $\hat O$ and the symmetry and transport properties of the system~\cite{Moudgalya2022}.
In the following study, we consider the infinite-temperature density autocorrelation function at site $j$:
\begin{equation}
C_j(t)=\langle\psi_{\mathrm{inf}}|[\hat n_j(t)-1/2][\hat n_j(0)-1/2]|\psi_{\mathrm{inf}}\rangle.
\end{equation}
At long times, $C_j(t)$ is larger than a lower bound~\cite{Sala2020,Moudgalya2022,Ghosh2023}
\begin{equation}
C_j^{(f)}=\frac{1}{\mathcal D} \sum_{\mathcal K}\frac{\left[\mathrm{Tr}(\hat P_{\mathcal K}\hat n_j\hat P_{\mathcal K}-\frac{1}{2})\right]^2}{\mathcal D_{\mathcal K}},
\end{equation}
where $\hat P_{\mathcal K}$ is the projector to the Krylov subspace $\mathcal K$ and the summation sums over all Krylov subspaces.
In the absence of HSF, $\mathcal K$ equals the symmetry sector and $C_j^{(f)}=0$ for the half-filling case.
In the presence of HSF, $C_j^{(f)}$ provides a finite lower bound for $C_j(t)$~\cite{Sala2020,Moudgalya2022,Ghosh2023}.

When the Hilbert space of a system is fragmented, one can redefine ergodicity within the Krylov subspaces~\cite{Moudgalya2021,Herviou2021}.
For an ergodic system without HSF, when evolving from an out-of-equilibrium state $|\psi_0\rangle$, the local observables should relax to the equilibrium values predicted by the ensemble of the symmetry sector.
In contrast, for a system with HSF, they should relax to the equilibrium values predicted by the ensemble of the Krylov subspace which the initial state belongs to, if the Krylov subspace is ergodic.
In the following, we study the evolution of the local densities $n_j(t)=\langle \hat n_j\rangle$.
For our system, $n_j$ will relax to 0.5 if there is no HSF.
Otherwise, it will relax to the value
\begin{equation}
n_j[{\mathcal K}]=\frac{1}{\mathcal D_{\mathcal K}}\mathrm{Tr}(\hat P_{\mathcal K}\hat n_j\hat P_{\mathcal K}),
\end{equation}
where $\mathcal K$ is the Krylov subspace which the initial state belongs to.

\subsection{Strong Hilbert space fragmentation without driving}\label{Sec.HSFu0}
\begin{figure}[!t]
\includegraphics[width=\columnwidth]{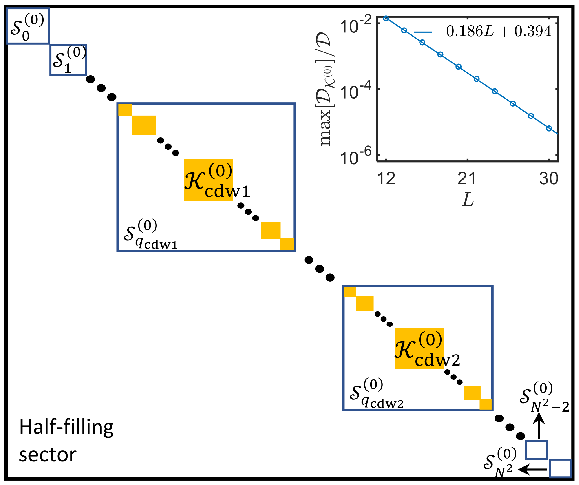}
  \caption{\label{Fig_Fragmentationu0}
  (a) Schematic of the allowed tunneling processes in the Hamiltonian $\hat H^{\mathrm{eff}}_{u=0}$, which conserve $\hat E$.
  (b) Schematic of the HSF in the half-filling sector under the action of $\hat H^{\mathrm{eff}}_{u=0}$.
  There are $N^2$ symmetry spaces $\mathcal S^{(0)}_q$ with good quantum number $E_q$, denoted by the empty blue squares.
  Apart from $\mathcal S^{(0)}_0, \mathcal S^{(0)}_1, \mathcal S^{(0)}_{N^2-2}$ and $\mathcal S^{(0)}_{N^2}$, the other symmetry spaces are further split into Krylov subspaces $\mathcal K^{(0)}_i$, denoted by the filled orange squares.
  The two largest Krylov subspaces $\mathcal K^{(0)}_{\mathrm{cdw1}}$ and $\mathcal K^{(0)}_{\mathrm{cdw2}}$ lie in the symmetry spaces $\mathcal S^{(0)}_{q_{\mathrm{cdw1}}}$ and $\mathcal S^{(0)}_{q_{\mathrm{cdw2}}}$ respectively.
  Inset in (b): the ratio between the dimension of the largest Krylov subspaces and that of the half-filling sector, as a function of the system size.
  The blue line is the fitting function.
  }
\end{figure}

When $u=0$, the Hamiltonian of the system reduces to $\hat H_0$ and the effective Hamiltonian up to first order (using Schrieffer-Wolff transformation~\cite{Bravyi2011}) in the large tilting limit is given by
\begin{equation}\label{Eq.Heffu0}
\hat H_{u=0}^{\mathrm{eff}}=J\sum_j\hat{\mathcal P}_{j-1,j+2}^{(0)}(\hatd c_j\hat c_{j+1}+\hatd c_{j+1}\hat c_j),
\end{equation}
where only the intrinsic resonance tunneling exists.
This effective Hamiltonian conserves the sum of the dipole moment and the number of pairs of occupied adjacent sites $\hat E=-\sum_jj\hat n_j+\sum_j\hat n_j\hat n_{j+1}$, see Fig.~\ref{Fig_Fragmentationu0}(a).

The action of Hamiltonian\eqref{Eq.Heffu0} fragments the Hilbert space beyond the simple conservation of $\hat E$ to Krylov subspaces $\mathcal K^{(0)}_i$, with the superscript denoting the splitting way under Hamiltonian\eqref{Eq.Heffu0} and the subscript denoting the different Krylov subspaces, see Fig.~\ref{Fig_Fragmentationu0}(b).
There are totally $N^2$ symmetry spaces $\mathcal S^{(0)}_q$ in the half-filling sector with good quantum numbers $E_q=E_0+q$, where $E_0=\frac{-3N(N-1)}{2}-1$ and $q=0,1,2,\cdots N^2-2$ and $N^2$.
Apart from the first and last two symmetry spaces (with dimensions 2, 3, 1, 1 for all system sizes), the Hilbert space in all the other symmetry spaces further fractures into disconnected Krylov subspaces.
There are two largest Krylov subspaces with the same dimension, labelled as $\mathcal K_{\mathrm{cdw1}}^{(0)}$ and $\mathcal K_{\mathrm{cdw2}}^{(0)}$, which respectively contain the charge density wave states $|\mathrm{CDW1}\rangle=|0 101\cdots01\rangle$ and $|\mathrm{CDW2}\rangle=|1 010\cdots1 0\rangle$, and lie in the symmetry spaces $\mathcal S_{q_{\mathrm{cdw1}}}^{(0)}$ with $E_{q_{\mathrm{cdw1}}}=-N^2$ and $\mathcal S_{q_{\mathrm{cdw1}}}^{(0)}$ with $E_{q_{\mathrm{cdw2}}}=-N(N-1)$.
In the inset of Fig.~\ref{Fig_Fragmentationu0}(b), we show $\max[\mathcal D_{\mathcal K^{(0)}}]/\mathcal D$, the dimension ratio between the largest Krylov subspace and the half-filling sector, as a function of the system size.
This ratio decays exponentially with $L$ and tends to 0 in the thermodynamic limit, which implies strong HSF regarding the half-filling sector.
We also calculate the dimension ratio between the largest Krylov subspace $\mathcal K_{\mathrm{cdw1}}^{(0)}$ [$\mathcal K_{\mathrm{cdw2}}^{(0)}$] and the corresponding symmetry space $\mathcal S_{q_{\mathrm{cdw1}}}^{(0)}$ [$\mathcal S_{q_{\mathrm{cdw2}}}^{(0)}$], which also decays exponentially with the system size and tends to 0 in the thermodynamic limit and implies strong HSF regarding $\mathcal S_{q_{\mathrm{cdw1}}}^{(0)}$ [$\mathcal S_{q_{\mathrm{cdw2}}}^{(0)}$].

\begin{figure}[!t]
\includegraphics[width=1\columnwidth]{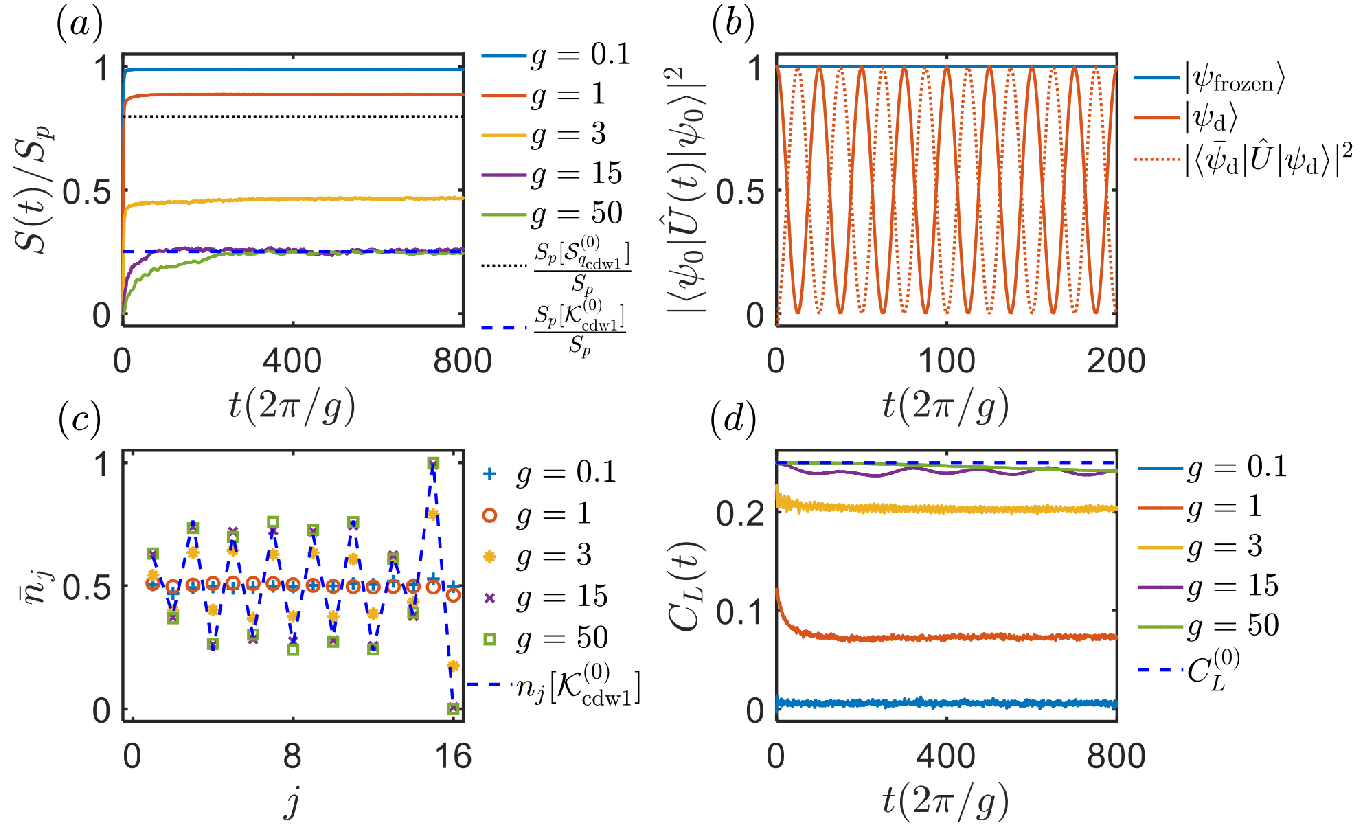}
  \caption{\label{Fig_SpObsAutoCs_u0}
  (a) The growth of the normalized von Neumann EE $S(t)/S_p$ for different values of $g$, averaged over all initial states (34 in total) in the largest Krylov subspace $\mathcal K_{\mathrm{cdw1}}^{(0)}$.
  The black dotted line and blue dashed lines denote $S_p[\mathcal S_{q_{\mathrm{cdw1}}}^{(0)}]/S_p$ and $S_p[\mathcal K_{\mathrm{cdw1}}^{(0)}]/S_p$, respectively.
  (b) At $g=50$, the fidelity dynamics from the frozen state $|\psi_{\mathrm{frozen}}\rangle=|1111111010000000\rangle$ (blue line), which lies in $\mathcal S^{(0)}_{N^2-2}$, and the domain state $|\psi_{\mathrm d}\rangle=|0000000011111111\rangle$ (red line), which lies in $\mathcal S^{(0)}_0$.
  Red dotted line shows the amplitude of state transfer between $|\psi_{\mathrm{d}}\rangle$ and $|\bar{\psi}_{\mathrm{d}}\rangle=|0000000101111111\rangle$.
  (c) The saturated density profile $\bar n_j$ for different values of $g$, averaged over time $t\in[700, 800]\frac{2\pi}{g}$ and all initial Fock states in $\mathcal K_{\mathrm{cdw1}}^{(0)}$.
  The blue dashed line denotes $n_j[\mathcal K_{\mathrm{cdw1}}^{(0)}]$.
  (d) The evolution of $C_L(t)$ for different values of $g$, starting from a random infinite-temperature state.
  The blue dashed line denotes the lower bound $C_L^{(0)}=0.25$, predicted by the HSF due to Hamiltonian\eqref{Eq.Heffu0}.
  In (a), (c) and (d), the data with $g=15$ and $50$ almost collapse with each other.
  The other parameters are $u=0$ and $L=16$ in all plots.
  The energies are scaled in units of $J$.
  }
\end{figure}

To demonstrate the HSF of $\hat H_0$ in the large tilting limit, we study the growth of EE starting from Fock states by the exact time evolution.
According to the random states\eqref{Eq.CanoRandomSts} in different spaces, we can calculate the Page value of EE, $S_p[\mathcal S^{(0)}_{q}]$, in the symmetry space $\mathcal S^{(0)}_q$, and $S_p[\mathcal K^{(0)}_i]$, in the Krylov subspace $\mathcal K^{(0)}_i$.
We calculate the growth of EE starting from all Fock states in $\mathcal K_{\mathrm{cdw1}}^{(0)}$, and show the average over all the initial states at different values of $g$ in Fig.~\ref{Fig_SpObsAutoCs_u0}(a).
For small values of $g$, $S(t)$ saturates to $S_p$ but not $S_p[{\mathcal S_{q_{\mathrm{cdw1}}}^{(0)}}]$, which indicates that the system is not fragmented and thermalizes within the half-filling sector.
For a larger $g$, $S(t)$ saturates to $S_p[\mathcal K_{\mathrm{cdw1}}^{(0)}]$, which demonstrates the splitting of Hilbert space under the action of $\hat H_{u=0}^{\mathrm{eff}}$.
The results in $\mathcal K_{\mathrm{cdw2}}^{(0)}$ are qualitatively the same, which are not shown here.

The HSF at large tilting strength is further evidenced by the saturated density profile.
In Fig.~\ref{Fig_SpObsAutoCs_u0}(c), we plot the saturated density profile $\bar n_j$ at different values of $g$.
It is obtained by averaging $n_j(t)$ over the times $t\in[700,800]\frac{2\pi}{g}$, during which $n_j(t)$ already saturates, and over all initial Fock states in $\mathcal K_{\mathrm{cdw1}}^{(0)}$.
For comparison, we also plot the density profile $n_j[\mathcal K^{(0)}_{\mathrm{cdw1}}]$, predicted by the infinite-temperature ensemble of the Krylov subspace $\mathcal K^{(0)}_{\mathrm{cdw1}}$.
For small values of $g$, $\bar n_j$ concentrate on 0.5 for all lattice sites, which is consistent with the result of thermalization within the half-filling sector.
For a larger $g$, the density profile is consistent with $n_j[{\mathcal K_{\mathrm{cdw1}}^{(0)}}]$, again implying that the Hilbert space splits according to $\hat H_{u=0}^{\mathrm{eff}}$.
The results in $\mathcal K_{\mathrm{cdw2}}^{(0)}$ are qualitatively the same, which are not shown here.

Apart from the large Krylov subspaces, there are numerous frozen states and small Krylov subspaces.
The frozen states correspond to the Fock states which are zero-energy eigenstates of $\hat H_{u=0}^{\mathrm{eff}}$.
Any Fock states are frozen under the action of $\hat H_{u=0}^{\mathrm{eff}}$, if they lack both the configurations $``\cdots0011\cdots"$ and $``\cdots0101\cdots"$, where the sequences $011$ and $101$ are away from the left particles by at least one site.
In Fig.~\ref{Fig_SpObsAutoCs_u0}(b), we plot the evolution of the fidelity $|\langle \psi_0|\hat U(t)|\psi_0\rangle|^2$ from a frozen state $|\psi_0\rangle=|\psi_{\mathrm{frozen}}\rangle$ at $g=50$.
The fidelity keeps near 1 for all the time considered, evidencing that the state stays to the initial state.
Furthermore, we study the dynamics in a Krylov subspace with two elements, which contains a state $|\psi_{\mathrm d}\rangle$ and its partner $|\bar{\psi}_{\mathrm{d}}\rangle=\hat H_{u=0}^{\mathrm{eff}}|\psi_{\mathrm d}\rangle$.
In Fig.~\ref{Fig_SpObsAutoCs_u0}(b), we plot the fidelity dynamics from $|\psi_0\rangle=|\psi_{\mathrm d}\rangle$ at $g=50$.
The fidelity oscillates periodically, and the state transfers almost perfectly between $|\psi_{\mathrm d}\rangle$ and $|\bar{\psi}_{\mathrm{d}}\rangle$, see the red solid and dotted lines in Fig.~\ref{Fig_SpObsAutoCs_u0}(b).
The oscillation period is $\tilde T_2=\pi/J$, which is consistent with the prediction by projecting the effective Hamiltonian $\hat H_{u=0}^{\mathrm{eff}}$ into the Krylov subspace.

At last, we study the autocorrelation function $C_{j=L}(t)$ starting from a random infinite-temperature state. The results make little difference for different infinite-temperature states.
The lower bound predicted by the present HSF is $C_L^{(0)}=0.25$.
In Fig.~\ref{Fig_SpObsAutoCs_u0}(d), we plot the evolution of $C_L(t)$ for different values of $g$.
For small values of $g$, the perturbation theory does not work, and $C_L(t)$ decays to zero quickly, which is the result of thermalization in the half-filling sector.
As $g$ increases, $C_L(t)$ saturates to finite values, and tends to the bound $C_L^{(0)}$.
All these results show that the half-filling sector is split strongly according to $\hat H^{\mathrm{eff}}_{u=0}$ in the large tilting limit and without driving.

\subsection{Strong Hilbert space fragmentation at the partial resonance frequency $\omega_1=g$}
\begin{figure}[b]
\includegraphics[width=\columnwidth]{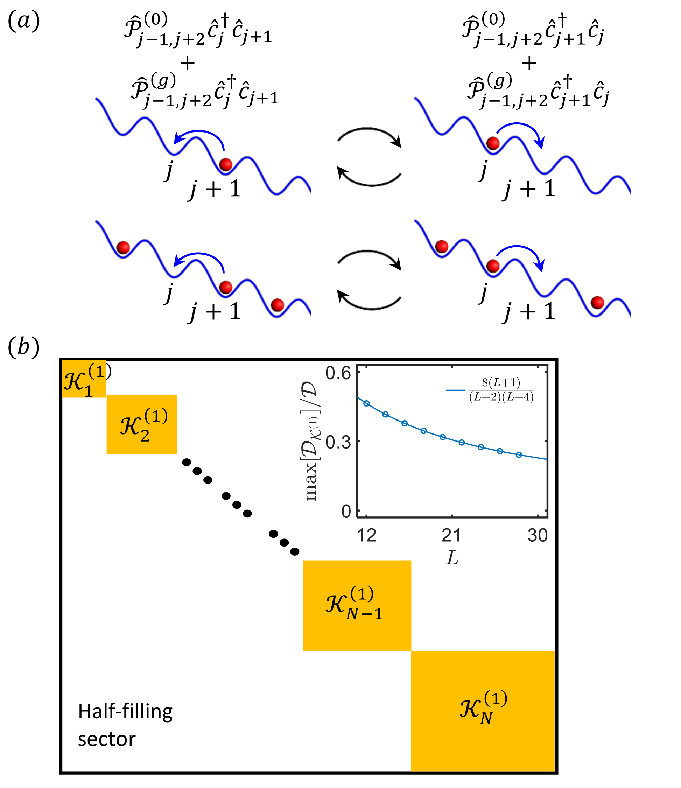}
  \caption{\label{Fig_FragmentationInter}
  (a) Schematic of the allowed tunneling processes in the Hamiltonian $\hat H^{\mathrm{eff}}_{\omega_1}$.
  (b) Schematic of the HSF in the half-filling sector under the action of $\hat H^{\mathrm{eff}}_{\omega_1}$.
  It is fragmented into $N$ Krylov subspaces $\mathcal K^{(1)}_i$ with the dimension increasing with $i$, denoted by the filled orange squares.
  Inset in (b): the ratio between the dimension of the largest Krylov subspace and that of the half-filling sector, as a function of the system size.
  The blue circles denote the numerical data and the blue line is the function $y=\frac{8(L+1)}{(L+2)(L+4)}$.
  }
\end{figure}

When the driving frequency $\omega=\omega_1$, the tunneling processes projected by $\hat P^{(0)}_{j-1,j+2}$ and $\hat P^{(g)}_{j-1,j+2}$ are allowed, while the one projected by $\hat P^{(2g)}_{j-1,j+2}$ vanishes, see schematic in Fig.~\ref{Fig_FragmentationInter}(a).
The first order effective Floquet Hamiltonian in this case reads
\begin{eqnarray}\label{Eq.HeffIntmediate}
\hat H_{\omega_1}^{\mathrm{eff}}&=&J\sum_j\hat{\mathcal P}_{j-1,j+2}^{(0)}(\hatd c_j\hat c_{j+1}+h.c.)\nonumber\\
&+&\frac{uJ}{2}\sum_j\hat{\mathcal P}_{j-1,j+2}^{(g)}(\hatd c_j\hat c_{j+1}+h.c.).
\end{eqnarray}

The additional tunneling term relaxes some kinetic constraints, thus changes the structure of the Hilbert space.
The Hamiltonian\eqref{Eq.HeffIntmediate} does not show obvious symmetry, and splits the half-filling sector into $N$ Krylov subspaces $\mathcal K_i^{(1)}$ $(i=1,\cdots,N)$, with the superscript denoting the splitting way under $\hat H_{\omega_1}^{\mathrm{eff}}$.
Each Krylov subspace $\mathcal K_i^{(1)}$ contains the state $|\psi_i\rangle=|(0\cdots0)^{i-1}(1\cdots1)^N(0\cdots0)^{N-i+1}\rangle$ with $(0\cdots0)^q$ and $(1\cdots1)^q$ denoting $q$ contiguous empty sites and occupied sites respectively.
The dimension of $\mathcal K_i^{(1)}$ increases with $i$ with $\mathcal K_1^{(1)}$ being the minimal subspace with dimension 1 for all system sizes, i.e., $|(1\cdots1)^N(0\cdots0)^N\rangle$ is a frozen state.
In Fig.~\ref{Fig_FragmentationInter}(b), we plot the dimension ratio $\max[\mathcal D_{\mathcal K^{(1)}}]/\mathcal D$ between the largest Krylov subspace and the half-filling sector as a function of the system size.
We numerically find that $\max[\mathcal D_{\mathcal K^{(1)}}]/\mathcal D=\frac{8(L+1)}{(L+2)(L+4)}$, with the analytical derivation deserving further study.
This ratio means that the largest Krylov subspace occupies a vanishing fraction of the half-filling sector in the thermodynamic limit and the system is strongly fragmented.

\begin{figure}[!htb]
\includegraphics[width=\columnwidth]{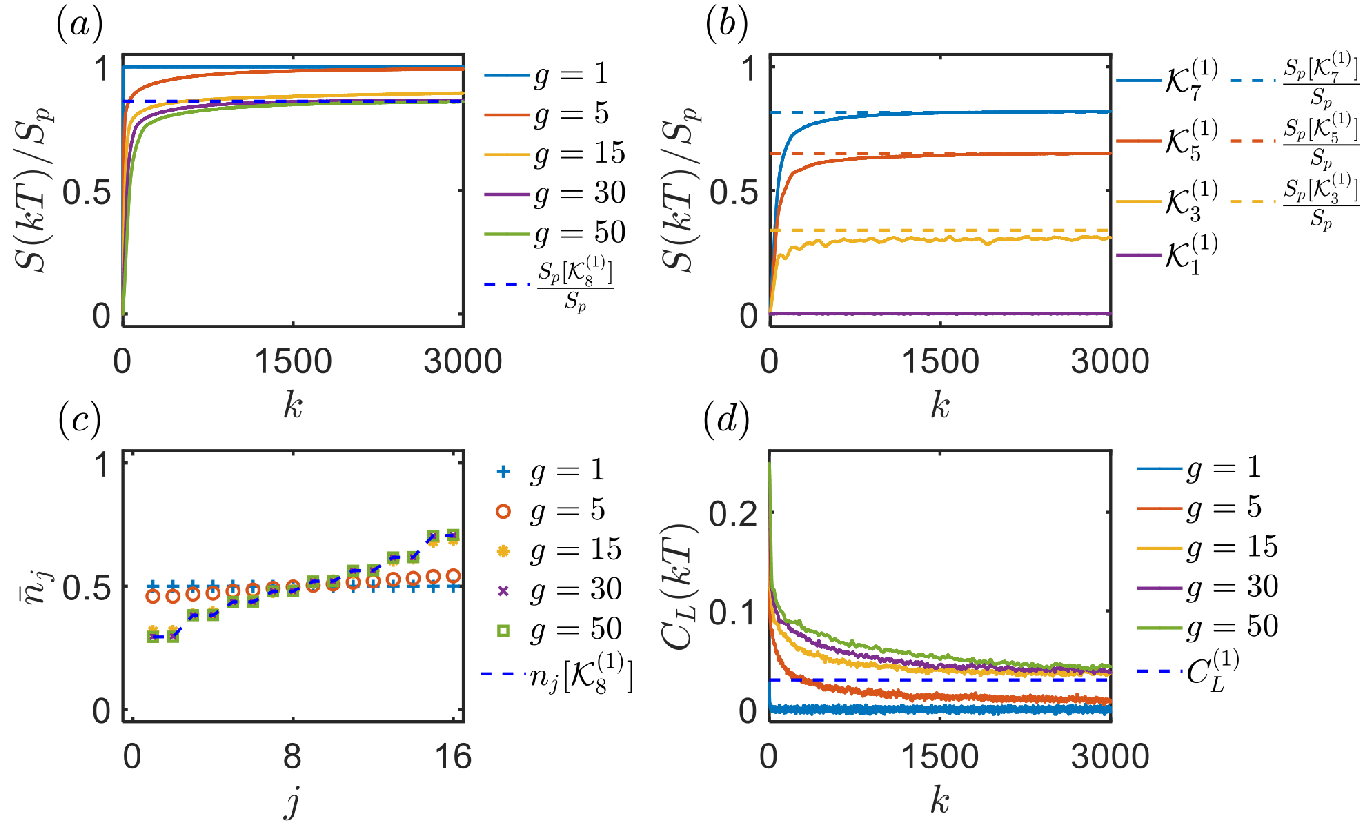}
  \caption{\label{Fig_SpObsAutoCs_Inter}
  (a) $S(kT)/S_p$ versus the driving cycles $k$ for different values of $g$, averaged over 50 randomly chosen Fock states in the largest Krylov subspace $\mathcal K_8^{(1)}$.
  The blue dashed line denotes $S_p[{\mathcal K_8^{(1)}}]/S_p$.
  (b) $S(kT)/S_p$ versus the driving cycles $k$ at $g=50$, averaged over 50 randomly chosen Fock states in $\mathcal K_7^{(1)}$ (blue line), $\mathcal K_5^{(5)}$ (red line) and $\mathcal K_3^{(1)}$ (yellow line), and starting from the Fock state in $\mathcal K_1^{(1)}$ (purple line).
  The dashed lines with different colors denote $S_p[\mathcal K_i^{(1)}]/S_p$ ($i=7,5,3$ and 1) of the corresponding Krylov subspaces $\mathcal K_i^{(1)}$.
  For $\mathcal K_1^{(1)}$, the Fock state is frozen and $S(kT)$ stays near 0.
  (c) The saturated density profile $\bar n_j$ for different values of $g$, averaged over the driving cycles $k\in[2900, 3000]$ and 50 randomly chosen Fock states in $\mathcal K_8^{(1)}$.
  The blue dashed line denotes $n_j[\mathcal K_8^{(1)}]$.
  (d) $C_L(kT)$ versus the driving cycles $k$ for different values of $g$, starting from the same infinite-temperature state in Fig.~\ref{Fig_SpObsAutoCs_u0}(d).
  The blue dashed line denotes the lower bound $C_L^{(1)}$ predicted by the HSF due to $\hat H_{\omega_1}^{\mathrm{eff}}$.
  In (a), (c) and (d), the numerical data with $g=30$ and $50$ almost collapse with each other.
  The other parameters are $u=1,\omega=g$ and $L=16$ in all plots.
  The energies are scaled in units of $J$.
  }
\end{figure}

To verify the strong HSF, we calculate the EE dynamics, the saturated density profile and the density autocorrelation function dynamics by the exact time evolution at $\omega=g$, $u=1$ and system size $L=16$.
In Fig.~\ref{Fig_SpObsAutoCs_Inter}(a), we plot $S(kT)/S_p$ versus the driving cycles $k$ at different values of $g$ for the largest Krylov subspace $\mathcal K^{(1)}_8$.
The EE are averaged over 50 initial random Fock states in $\mathcal K^{(1)}_8$.
For small values of $g$, $S(kT)$ saturates to $S_p$ rapidly.
As $g$ increases, $S(kT)$ starts to saturate to smaller values, and reaches $S_p[\mathcal K^{(1)}_8]$ finally.
At a large value of $g$, the EE for Fock states in the other Krylov subspaces also saturates to the Page value of the corresponding Krylov subspaces.
In Fig.~\ref{Fig_SpObsAutoCs_Inter}(b), we plot $S(kT)/S_p$ at $g=50$ in Krylov subspaces $\mathcal K^{(1)}_7$, $\mathcal K^{(1)}_5$, $\mathcal K^{(1)}_3$ and $\mathcal K^{(1)}_1$.
For $\mathcal K^{(1)}_{7,5,3}$, the EE are averaged over 50 initial random Fock states, and they saturate to values $S_p[\mathcal K^{(1)}_{7,5,3}]$ respectively.
For $\mathcal K^{(1)}_1$, the EE keeps near 0, which is consistent with the frozen state prediction.
Fig.~\ref{Fig_SpObsAutoCs_Inter}(c) shows the saturated density profile $\bar n_j$ at different values of $g$, for the same initial states as that in Fig.~\ref{Fig_SpObsAutoCs_Inter}(a).
It is obtained by averaging $n_j(kT)$ over the driving cycles $k\in[2900,3000]$, during which $n_j(kT)$ already saturates, and over all the 50 initial states.
For small values of $g$, $\bar n_j$ shows a uniform profile around 0.5.
For large values of $g$, $\bar n_j$ is consistent with $n_j[\mathcal K^{(1)}_8]$.
At last, Fig.~\ref{Fig_SpObsAutoCs_Inter}(d) shows the evolution of $C_L(kT)$ for different values of $g$, starting from the same infinite-temperature state as that in Fig.~\ref{Fig_SpObsAutoCs_u0}(d).
The results make little difference for different infinite-temperature states.
For small $g$, $C_L(kT)$ decays to zero quickly.
For a large $g$, $C_L(kT)$ saturates to $C_L^{(1)}$ predicted by the present HSF.
All these features demonstrate that the half-filling sector is split strongly according to $\hat H^{\mathrm{eff}}_{\omega_1}$ at $\omega=\omega_1$ in the large tilting limit.

\subsection{Strong Hilbert space fragmentation at the partial resonance frequencies $\omega_2=2g$}
\begin{figure}[!htb]
\includegraphics[width=\columnwidth]{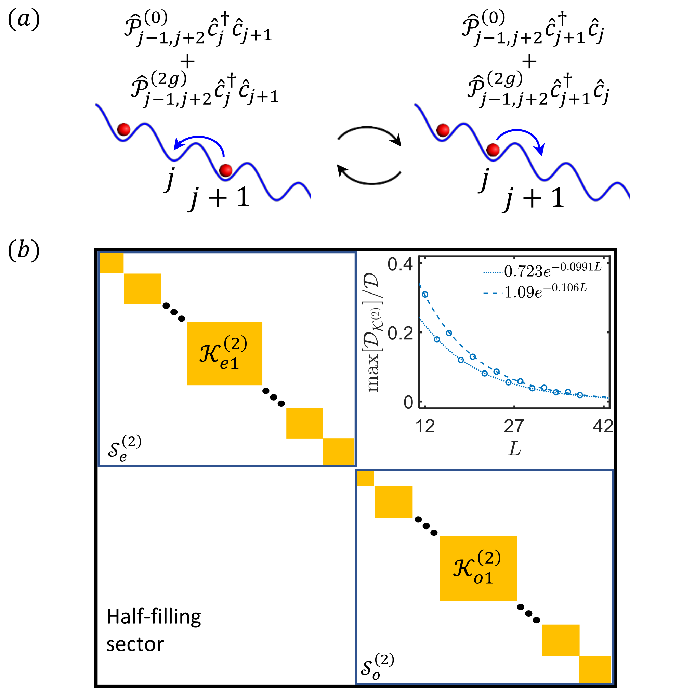}
  \caption{\label{Fig_FragmentationSHSF}
  (a) Schematic of the allowed tunneling processes in the Hamiltonian $\hat H^{\mathrm{eff}}_{\omega_2}$, which conserve the parity of $\hat E$.
  (b) Schematic of the HSF in the half-filling sector under the action of $\hat H^{\mathrm{eff}}_{\omega_2}$.
  The even and odd parity symmetry spaces $\mathcal S^{(2)}_{e}$ and $\mathcal S^{(2)}_{o}$ are denoted by the empty blue squares.
  $\mathcal S^{(2)}_{e}$ and $\mathcal S^{(2)}_{o}$ are further fragmented into Krylov subspaces $\mathcal K^{(2)}_i$, denoted by the filled orange squares.
  Inset in (b): the ratio between the dimension of the largest Krylov subspace and that of the half-filling sector, as a function of the system size.
  The blue circles denote the numerical data.
  The blue dashed and dotted lines denote the fitting functions for quadruple and non-quadruple lattice sites, respectively.
  }
\end{figure}

When the driving frequency $\omega=\omega_2$, the tunneling processes projected by $\hat P^{(0)}$ and $\hat P^{(2g)}$ are resonant.
The tunneling process projected by $\hat P^{(g)}$ is off-resonant, but yet does not vanish.
The first order effective Floquet Hamiltonian in this case reads
\begin{eqnarray}
\hat H_{\omega_2}^{'\mathrm{eff}}&=&\frac{g}{2}\left[1-(-1)^{\sum_j\left(-j\hat n_j+\hat n_j\hat n_{j+1}\right)}\right]\nonumber\\
&+&J\sum_j\hat{\mathcal P}_{j-1,j+2}^{(0)}(\hatd c_j\hat c_{j+1}+\hatd c_{j+1}\hat c_j)\nonumber\\
&-&\frac{2iJ(3-u)}{3\pi}\sum_j\hat{\mathcal P}_{j-1,j+2}^{(g)}(\hatd c_{j+1}\hat c_j-\hatd c_j\hat c_{j+1})\nonumber\\
&+&\frac{uJ}{2}\sum_j\hat{\mathcal P}_{j-1,j+2}^{(2g)}(\hatd c_j\hat c_{j+1}+\hatd c_{j+1}\hat c_j).
\end{eqnarray}
At first sight, all the kinetic constraints are released now.
However, due to the effective on-site potential, the tunneling process projected by $\hat{\mathcal P}^{(g)}$ is biased by energy $g$, which is much larger than the tunneling amplitude $2J(3-u)/3\pi$.
Thus, to first order in $J/g$, we can neglect this tunneling process and reduce the effective Hamiltonian to
\begin{eqnarray}\label{Eq.HeffStrong}
\hat H_{\omega_2}^{\mathrm{eff}}&=&J\sum_j\hat{\mathcal P}_{j-1,j+2}^{(0)}(\hatd c_j\hat c_{j+1}+\hatd c_{j+1}\hat c_j)\nonumber\\
&+&\frac{uJ}{2}\sum_j\hat{\mathcal P}_{j-1,j+2}^{(2g)}(\hatd c_j\hat c_{j+1}+\hatd c_{j+1}\hat c_j),
\end{eqnarray}
where the effective on-site potential has been dropped, since the two tunneling processes conserve the parity $\hat{\mathcal P}_E=(-1)^{\hat E}$ of $\hat E=\sum_j\left(-j\hat n_j+\hat n_j\hat n_{j+1}\right)$, see Fig.~\ref{Fig_FragmentationSHSF}(a).

The action of $\hat H^{\mathrm{eff}}_{\omega_2}$ splits the half-filling sector into Krylov subspaces $\mathcal K^{(2)}_i$ beyond the conservation of $\hat{\mathcal P}_E$, where the superscript of $\mathcal K^{(2)}_i$ denotes the splitting way under $\hat H_{\omega_2}^{\mathrm{eff}}$ and the subscript denotes the different Krylov subspaces, see Fig.~\ref{Fig_FragmentationSHSF}(b).
It should be noted that the two allowed processes are related by a spatial-reflection transformation, i.e. $\hat R [\hat{\mathcal P}^{(0)}_{j-1,j+2}(\hatd c_j\hat c_{j+1}+h.c.)]\hat R^{-1}=\hat{\mathcal P}^{(2g)}_{L-j-1,L-j+2}(\hatd c_{L-j+1}\hat c_{L-j}+h.c.)$, where the spatial-reflection operator $\hat R$ is defined by $\hat R|n_0n_1\cdots n_{L-2}n_{L-1}\rangle=|n_{L-1}n_{L-2}\cdots n_1n_0\rangle$.
This leads to an odd-even effect of the particle number on the splitting of the Hilbert space (see Appendix~\ref{AppA} for details).
When the particle number $N$ is odd, the Krylov subspaces in the even- and odd-parity symmetry space $\mathcal S^{(2)}_e$ and $\mathcal S^{(2)}_o$ are in one-to-one correspondence through the spatial-reflection transformation.
Thus, there are two largest Krylov subspace with the same dimension, labelled as $\mathcal K^{(2)}_{e1}$ and $\mathcal K^{(2)}_{o1}$, which lie in $\mathcal S^{(2)}_e$ and $\mathcal S^{(2)}_o$ respectively.
When $N$ is even, there exists reflection-invariant Krylov subspace and the largest Krylov subspace lies in $\mathcal S^{(2)}_e$.
So, the scalings of the dimension ratio between the largest Krylov subspace and the half-filling sector is different for quadruple and non-quadruple lattice sites, and we need two functions to fit them.
In the inset of Fig.~\ref{Fig_FragmentationSHSF}(b), we plot this ratio as a function of the system size.
It shows exponential decay with the system size and vanishes in the thermodynamic limit, which implies strong HSF regarding the half-filling sector.
We also calculate the dimension ratio between the largest Krylov subspace $\mathcal K_{e1}^{(2)}$ [$\mathcal K_{o1}^{(2)}$] within the even (odd)-parity symmetry space $\mathcal S_e^{(2)}$ [$\mathcal S_o^{(2)}$] and that of $\mathcal S_e^{(2)}$ [$\mathcal S_o^{(2)}$].
They also decays exponentially with the system size and tend to 0 in the thermodynamic limit, and implies strong HSF regarding $\mathcal S_e^{(2)}$ and $\mathcal S_o^{(2)}$.

\begin{figure}[!htb]
\includegraphics[width=\columnwidth]{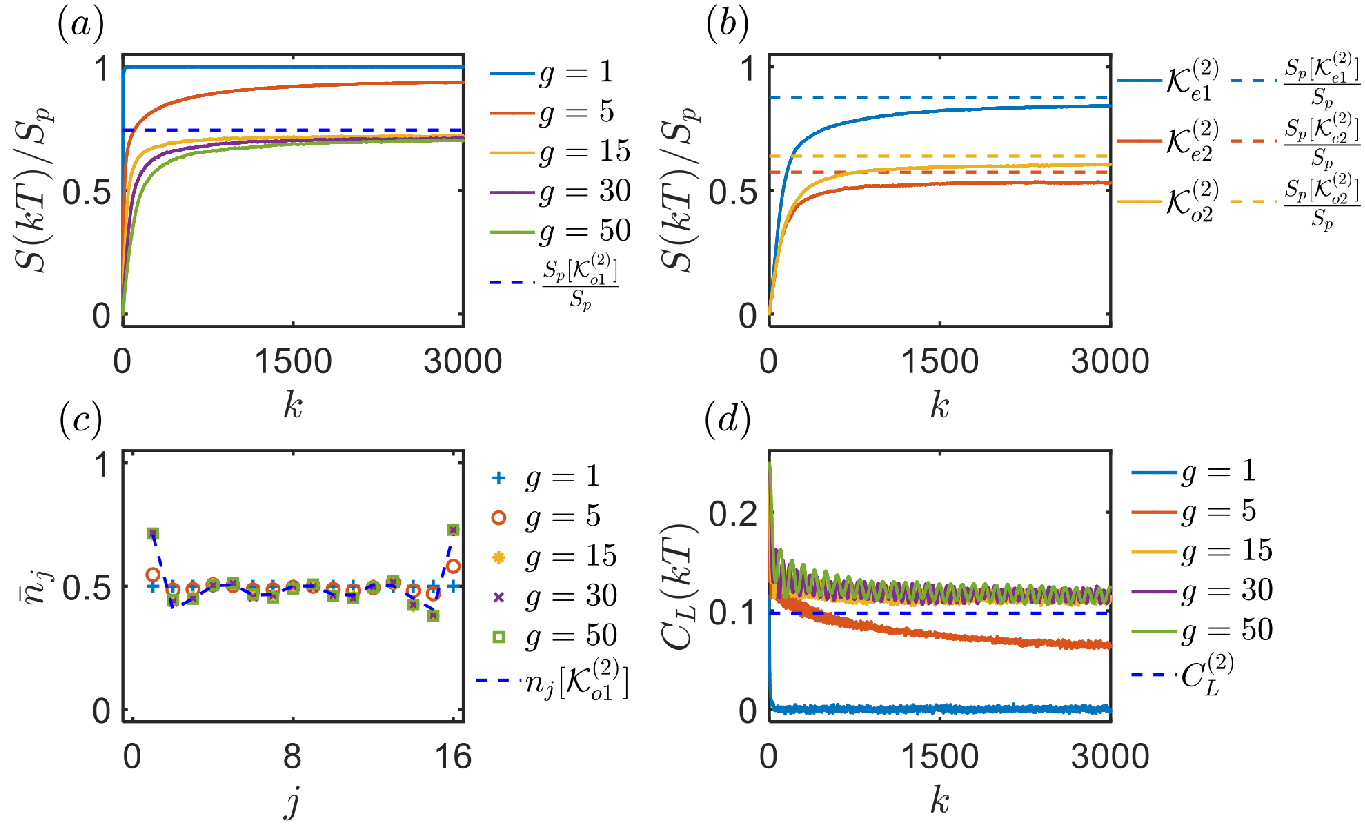}
  \caption{\label{Fig_SpObsAutoCs_Strong}
  (a) $S(kT)/S_p$ versus the driving cycles $k$ for different values of $g$, averaged over 50 randomly chosen Fock states in $\mathcal K^{(2)}_{o1}$.
  The blue dashed line denotes $S_p[\mathcal K^{(2)}_{o1}]/S_p$.
  (b) $S(kT)/S_p$ versus the driving cycles $k$ at $g=50$, averaged over 50 randomly chosen Fock states in $\mathcal K^{(2)}_{e1}$ (blue line), $\mathcal K^{(2)}_{e2}$ (red line) and $\mathcal K^{(2)}_{o2}$ (yellow line).
  The dashed lines with different colors denote the normalized Page value of EE of the corresponding Krylov subspaces.
  (c) The saturated density profile $\bar n_j$ for different values of $g$, averaged over the driving cycles $k\in[2900,3000]$ and 50 randomly chosen Fock states in $\mathcal K^{(2)}_{o1}$.
  The blue dashed line denotes $n_j[\mathcal K^{(2)}_{o1}]$.
  (d) $C_L(kT)$ versus $k$ for different values of $g$, starting from the same infinite-temperature state in Fig.~\ref{Fig_SpObsAutoCs_u0}(d).
  The blue dashed line denotes the lower bound $C_L^{(2)}$ predicted by the HSF due to $\hat H^{\mathrm{eff}}_{\omega_2}$.
  In (a), (c) and (d), the numerical data with $g=15,30$ and $50$ almost collapse with each other.
  The other parameters are $u=1,\omega=2g$ and $L=16$ in all plots.
  The energies are scaled in units of $J$.
  }
\end{figure}

To verify this strong HSF, we calculate the EE dynamics, the saturated density profile and the density autocorrelation function dynamics by the exact time evolution at $\omega=2g$ and $u=1$.
In Fig.~\ref{Fig_SpObsAutoCs_Strong}(a), we plot $S(kT)/S_p$ versus the driving cycles $k$ at different values of $g$ for the largest Krylov subspace $\mathcal K^{(2)}_{o1}$ within the odd-parity space.
The EE are averaged over 50 initial random Fock states in $\mathcal K^{(2)}_{o1}$.
For small values of $g$, $S(kT)$ saturates to $S_p$ rapidly.
As $g$ increases, $S(kT)$ starts to saturate to smaller values, and reaches $S_p[\mathcal K^{(2)}_{o1}]$ finnaly.
At a large value of $g$, the EE for Fock states in other Krylov subspaces also saturates to the Page value of the corresponding Krylov subspaces.
In Fig.~\ref{Fig_SpObsAutoCs_Strong}(b), we plot $S(kT)/S_p$ at $g=50$ within the largest and second largest Krylov subspace $\mathcal K^{(2)}_{e1}$ and $\mathcal K^{(2)}_{e2}$ within the even-parity space, and the second largest Krylov subspace $\mathcal K^{(2)}_{o2}$ within the odd-parity space.
For all the Krylov subspaces, the EE are calculated by averaging 50 initial random Fock states, and they saturate to the Page values of the corresponding Krylov subspaces.
Fig.~\ref{Fig_SpObsAutoCs_Strong}(c) shows the saturated density profile $\bar n_j$ at different values of $g$, for the same initial states as that in Fig.~\ref{Fig_SpObsAutoCs_Strong}(a).
It is obtained by averaging $n_j(kT)$ over the driving cycles $k\in[2900,3000]$, during which $n_j(kT)$ already saturates, and over all the initial states.
For small values of $g$, $\bar n_j$ shows a uniform profile around 0.5.
For large values of $g$, $\bar n_j$ is consistent with $n_j[\mathcal K^{(2)}_{o1}]$.
At last, Fig.~\ref{Fig_SpObsAutoCs_Strong}(d) shows the evolution of $C_L(kT)$ for different values of $g$, starting from the same infinite-temperature state as that in Fig.~\ref{Fig_SpObsAutoCs_u0}(d).
The results make little difference for different infinite-temperature states.
For small $g$, $C_L(kT)$ decays to zero quickly.
For a large $g$, $C_L(kT)$ saturates to $C_L^{(2)}$ predicted by the present strong HSF.
All these features demonstrate the half-filling sector is split strongly under $\hat H^{\mathrm{eff}}_{\omega_2}$ at $\omega=\omega_2$ in the large tilting limit.

\subsection{Strong Hilbert space fragmentation at non-resonant frequencies}
\begin{figure}[!t]
\includegraphics[width=\columnwidth]{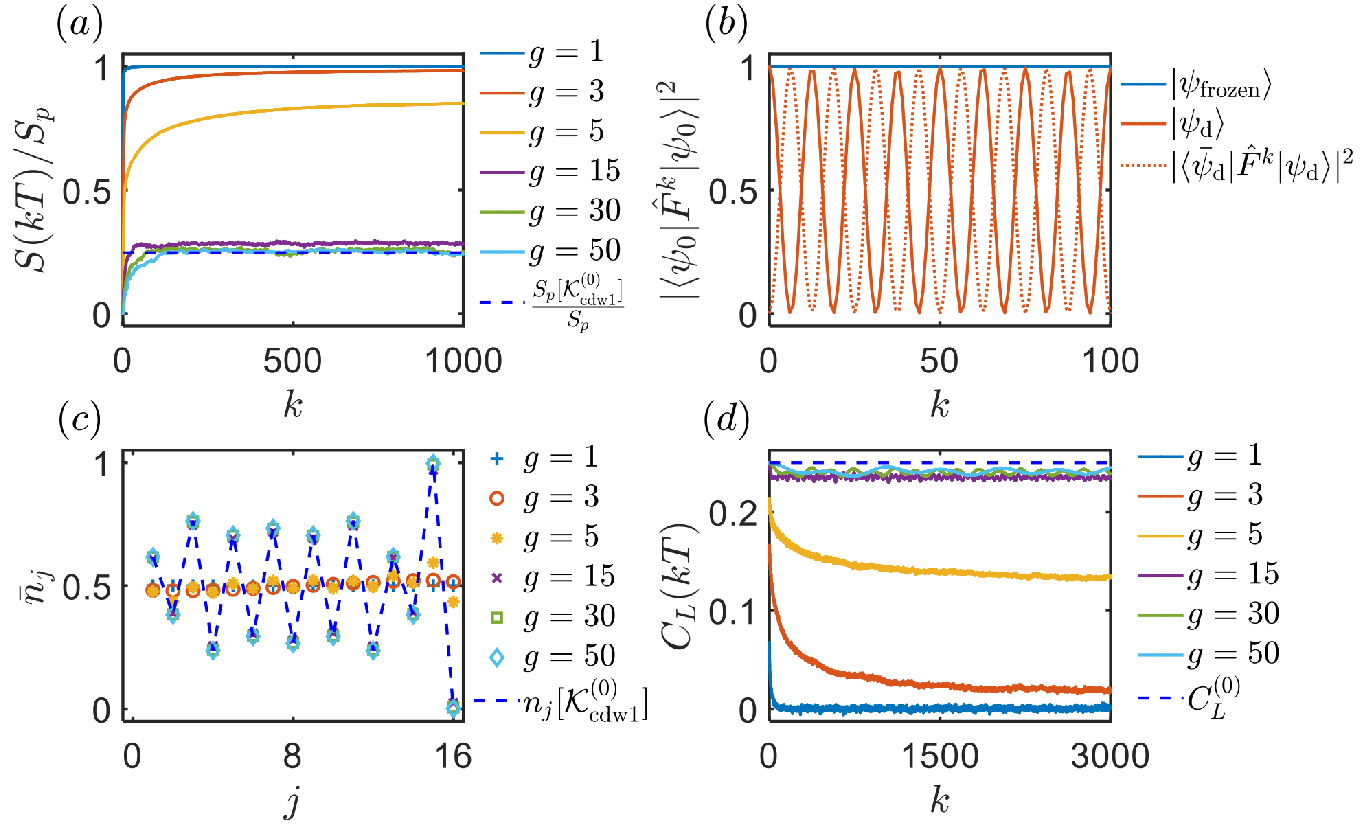}
  \caption{\label{Fig_SpObsAutoCs_omega2g}
  (a) The growth of the averaged and normalized von Neumann EE $S(kT)/S_p$ for different values of $g$.
  (b) The fidelity dynamics at $g=50$, from the frozen state $|\psi_{\mathrm{frozen}}\rangle$ and the domain state $|\psi_{\mathrm{d}}\rangle$.
  Red dotted line shows the amplitude of state transfer between $|\psi_{\mathrm{d}}\rangle$ and $|\bar{\psi}_{\mathrm{d}}\rangle$.
  (c) The saturated density profile $\bar n_j$ for different values of $g$, averaged over the driving cylcles $k\in[900, 1000]$ and all the initial states.
  (d) The evolution of $C_L(kT)$ for different values of $g$.
  The initial states in (a)-(d) are the same as that in Fig.~\ref{Fig_SpObsAutoCs_u0}(a)-(d) respectively.
  In (a), (c) and (d), the data with $g=15,30$ and $50$ almost collapse with each other.
  The other parameters are $u=1$, $\omega=g/2$ and $L=16$ in all plots.
  The energies are scaled in units of $J$.
  }
\end{figure}

When the driving frequency $\omega$ is away from the partial resonant values $\omega_1$ and $\omega_2$, the tunneling strength of the processes projected by $\hat{\mathcal P}^{(g)}$ and $\hat{\mathcal P}^{(2g)}$ is much smaller than the effective potential bias, and the two tunneling processes can be neglected.
One then expects the same HSF as the case without driving.
Specifically, when $\omega=g/q$ with $q$ being integers larger than 1, the tunneling strength of the two processes equals 0.
We demonstrate the strong HSF at the non-resonant frequencies by studying the EE dynamics, the fidelity dynamics, the saturated density profile and the density autocorrelation function dynamics through the exact time evolution, see Fig.~\ref{Fig_SpObsAutoCs_omega2g} for $\omega=g/2$ and $u=1$.
It should be noted that for the two-dimensional Krylov subspace, the period of fidelity oscillation in units of the driving period $T$ is predicted as $\tilde T_2/T=\omega/2J$.
All the observables show qualitatively similar behaviour as that in Fig.~\ref{Fig_SpObsAutoCs_u0}, thus demonstrating that in the large tilting limit and at non-resonant frequencies, the Hilbert space is split in the same way as that in Sec.~\ref{Sec.HSFu0}.

\subsection{Tuning the Hilbert space fragmentation by controlling the driving frequency at large tilting strength}
Now, we have established that the Hilbert space will fragment in three different ways when tuning the driving frequency in the large tilting limit.
To make it clearer, we study the evolution of $S(kT)$ from one Fock state $|\psi_0\rangle$ and of $C_L(kT)$ from one infinite-temperature state.

\begin{figure}[!htb]
\includegraphics[width=1\columnwidth]{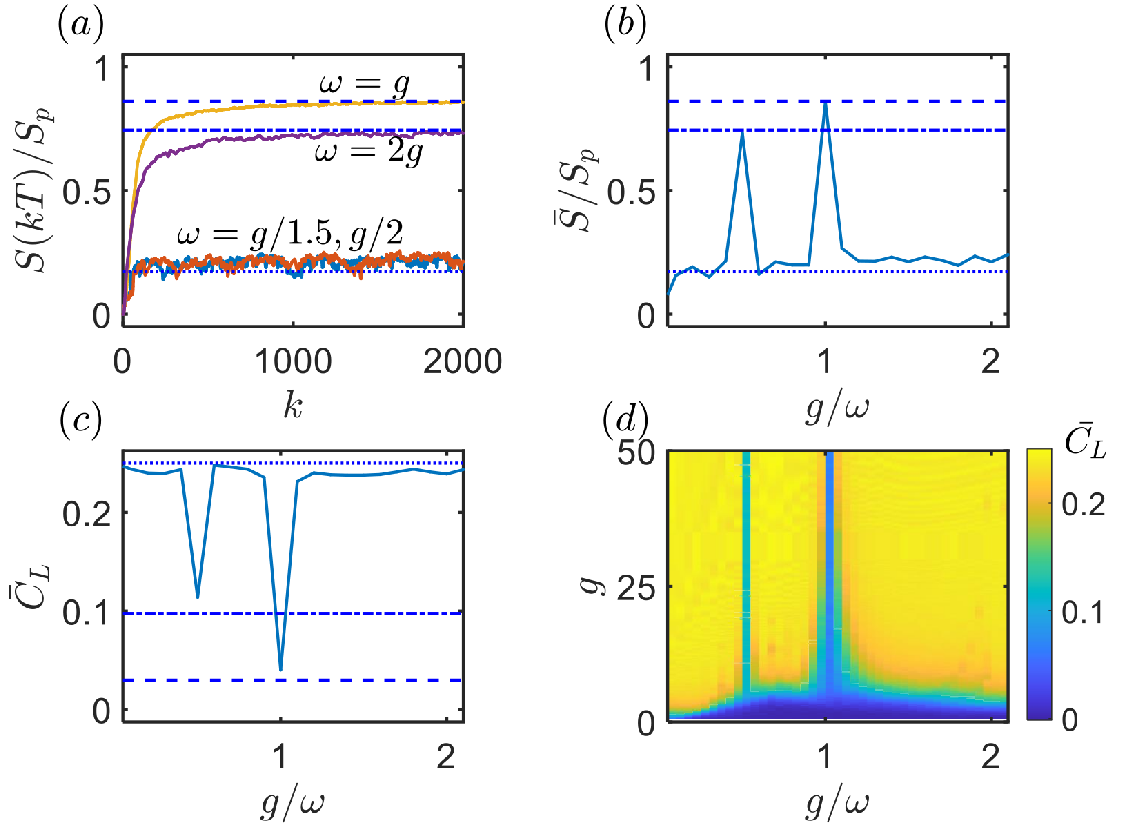}
  \caption{\label{Fig_SpAutoCs_omega}
  (a) $S(kT)/S_p$ versus the driving cycles $k$ for different driving frequency $\omega=g/2$ (blue line), $g/1.5$ (red line), $g$ (yellow line) and $2g$ (purple line).
  The initial state is $|\psi_0\rangle=|1010011001101001\rangle$.
  (b) $\bar S/S_p$ versus $g/\omega$, with the same initial state in (a).
  In (a) and (b), the blue dotted, dashed and dot-dashed lines denote the three normalized Page values $S_p[\mathcal K_{\psi_0}^{(0)}]/S_p$, $S_p[\mathcal K_{\psi_0}^{(1)}]/S_p$ and $S_p[\mathcal K_{\psi_0}^{(2)}]/S_p$ respectively.
  (c) $\bar C_L$ versus $g/\omega$, starting from the same infinite-temperature state in Fig.~\ref{Fig_SpObsAutoCs_u0}(d), ~\ref{Fig_SpObsAutoCs_Inter}(d) and ~\ref{Fig_SpObsAutoCs_Strong}(d).
  In (a)-(c), the other parameters are $u=1$, $g=50$ and $L=16$.
  (d) $\bar C_L$ as a function of $(g,g/\omega)$ at $u=1$ and system size $L=12$.
  In all plots, the energies are scaled in units of $J$.
  }
\end{figure}

In Fig.~\ref{Fig_SpAutoCs_omega}(a), we plot the growth of the normalized EE from the initial state $|\psi_0\rangle=|1010011001101001\rangle$, for $\omega=g/2, g/1.5, 2g$ and $g$, $g=50$ and $u=1$.
The initial state is chosen in some rand way, but requiring that the Page values of EE of the three Krylov subspaces $\mathcal K^{(0)}_{\psi_0}$, $\mathcal K^{(1)}_{\psi_0}$ and $\mathcal K^{(2)}_{\psi_0}$ to which $|\psi_0\rangle$ belongs make sufficient difference for better illustration.
For $\omega=g$ and $2g$, the EE saturates to $\mathcal S_p[\mathcal K^{(1)}_{\psi_0}]$ and $\mathcal S_p[\mathcal K^{(2)}_{\psi_0}]$ respectively.
For the other two frequencies, the EE saturates to $\mathcal S_p[\mathcal K^{(0)}_{\psi_0}]$.
In Fig.~\ref{Fig_SpAutoCs_omega}(b), we plot the normalized saturated EE $\bar{S}/S_p$ versus $g/\omega$ for $|\psi_0\rangle$ at $g=50$ and $u=1$.
$\bar S$ is calculated by averaging $S(kT)$ over the driving cycles $k\in[1900,2000]$.
We see two clear peaks at $\omega=\omega_1$ and $\omega_2$, where the saturated EE is consistent with $\mathcal S_p[\mathcal K^{(1)}_{\psi_0}]$ and $\mathcal S_p[\mathcal K^{(2)}_{\psi_0}]$ respectively.
At other frequencies, the saturated EE is consistent with $\mathcal S_p[\mathcal K^{(0)}_{\psi_0}]$.

The saturated EE depends on the initial state strongly.
The saturated autocorrelation function barely depends on the specific form of the initial infinite-temperature state, thus serves as a better criterion to distinguish the HSF.
In Fig.~\ref{Fig_SpAutoCs_omega}(c), we plot the saturated density autocorrelation function $\bar C_L$ versus $g/\omega$ from the infinite-temperature state in Fig.~\ref{Fig_SpObsAutoCs_u0}(d), at $g=50$ and $u=1$.
$\bar C_L$ is obtained by averaging $C_L(kT)$ over the driving cycles $k\in[5900,6000]$.
It shows two dips at $\omega=\omega_1$ and $\omega_2$, which is consistent with the predicted value $C_L^{(1)}$ and $C_L^{(2)}$ respectively.
For other values of $\omega$, $\bar C_L$ is consistent with $C_L^{(0)}$.

Finally, in Fig.~\ref{Fig_SpAutoCs_omega}(d), we plot a phase diagram on the parameter plane $(g,g/\omega)$ at $u=1$ and $L=12$, which shows the thermal and three kinds of HSF features of $\bar C_L$.
For $\omega=\omega_1$ and $\omega_2$, $\bar C_L$ respectively crossovers from 0 to $C_L^{(1)}$ and $C_L^{(2)}$ as $g$ increases, while for other values of $\omega$, it crossovers to $C_L^{(0)}$.
All these features demonstrate that one can tune the HSF by controlling the driving frequency when the tilting strength is sufficiently strong.

\section{Summary}\label{Sec4}
In conclusion, we have proposed a novel scheme to tune the HSF within the half-filling sector, through Floquet engineering of constrained hoppings in strongly tilted lattices.
In the absence of driving, the system is kinetically constrained and stays in a strong fragmented phase.
In the presence of driving, we engineer two effective Hamiltonians which release some kinetic constraints in different ways, by utilizing the partial resonance between the driving and the system.
We give the two partial resonance frequencies analytically by using the time-dependent perturbation theory in Floquet system.
The releasing of the kinetic constraints changes the structure of the Hilbert space and leads to another two kinds of strong Hilbert space fragmentation.
At the non-resonant frequencies, the system remains in the strong fragmented phase without driving.
We demonstrate the three kinds of Hilbert space fragmentation by studying the EE dynamics, the saturated density profile and the density autocorrelation functions.
Our results provide an efficient way to tune the Hilbert space fragmentation by controlling the driving frequency of particle tunneling in tilted and interacting lattice systems.
The possibility to control the hopping channels through tuning the resonance between the driving and the system provides a new opportunity to study the relation between kinetic constraints and other ergodicity breaking phenomenon, such as the quantum many body scars~\cite{Serbyn2021,Guo2023}.

\acknowledgments
This study is supported by the National Natural Science Foundation of China (Grants No. 12025509, No. 12305048) and the National Key Research and Development Program of China (Grant No. 2022YFA1404104). Y.K. is supported by the National Natural Science Foundation of China (Grant No. 12275365) and the Natural Science Foundation of Guangdong (Grant No. 2023A1515012099).

\appendix
\setcounter{figure}{0}
\renewcommand{\thefigure}{A\arabic{figure}}

\section{Proposal for the cosine driving protocol}\label{AppA}
We discuss a scheme to realize our model $\hat H(t)=\hat H_J(t)+\hat H_U+\hat H_g$ in Rydberg atom platform, where $\hat H_J(t)=J[1+u\cos(\omega t)]\sum_j(\hatd c_j\hat c_{j+1}+h.c.)$, $\hat H_U=U\sum_j\hat n_j\hat n_{j+1}$ and $\hat H_g=-g\sum_jj\hat n_j$.

We consider a one-dimensional chain of $L$ Rydberg atoms, with lattice constant $a$ and each trapped in optical tweezers~\cite{Barredo2016,Endres2016,Browaeys2016,Barredo2018,Browaeys2020}.
Two Rydberg states are chosen to simulate the empty state and the fermions occupied state with $|0\rangle=|nS\rangle$ and $|1\rangle=|n'P\rangle$, where $|nS\rangle$ and $|n'P\rangle$ are the two Rydberg states with principal quantum numbers $n$ and $n'$ ($n\simeq n'$) and angular momentum $S$ and $P$ respectively (we note that $n$ labels the principal quantum number in this appendix and the fermions number in the main text).
The Rydberg atom chain is subjected to a gradient magnetic field which causes an effective Zeeman splitting $M_j$ between the two Rydberg states with $M_{j}-M_{j+1}=\tilde g$, as described the Hamiltonian $\hat H_{\tilde g}=-\tilde g\sum_jj(|n'P\rangle\langle n'P|)_j$.

\begin{figure}[t]
\includegraphics[width=1\columnwidth]{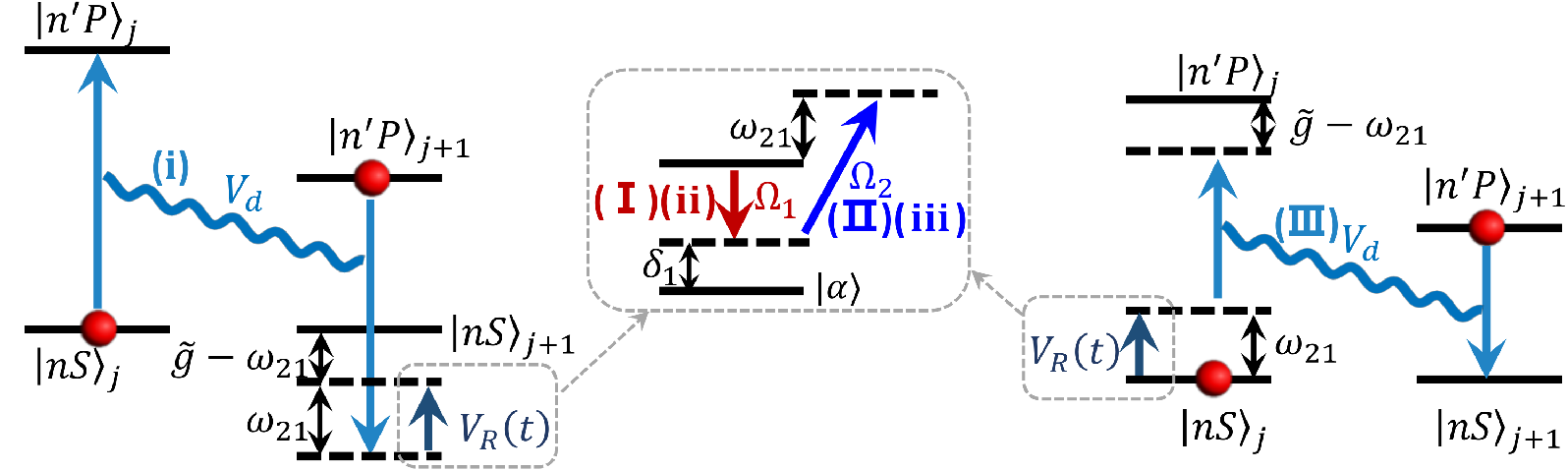}
  \caption{\label{Fig_LDD}
  The two Raman-assisted three-order transition processes coupling $|nS\rangle_j|n'P\rangle_{j+1}$ to $|n'P\rangle_j|nS\rangle_{j+1}$.
  The left energy level scheme denotes the three-order transition process $|nS\rangle_j|n'P\rangle_{j+1}\to|n'P\rangle_j|nS\rangle_{j+1}\to |n'P\rangle_j|\alpha\rangle_{j+1}\to|n'P\rangle_j|nS\rangle_{j+1}$, in which the first process (i) is the bare exchange interaction and the second (ii) and third processes (iii) induces a two-photon detuning $\omega_{21}=\omega_{\mathrm{R}_2}-\omega_{\mathrm{R}_1}$.
  The right energy level scheme denotes the three-order transition process $|nS\rangle_j|n'P\rangle_{j+1}\to|\alpha\rangle_j|n'P\rangle_{j+1}\to |nS\rangle_j|n'P\rangle_{j+1}\to|n'P\rangle_j|nS\rangle_{j+1}$, in which the first and second processes (I) and (II) induces the two-photon detuning $\omega_{21}$ and the third process (III) is the exchange interaction.
  }
\end{figure}

Utilizing the gradient magnetic field and a laser-assisted dipole-dipole (DD) interaction between Rydberg atoms~\cite{Yang2022}, we can realize the term $\hat H_g+\hat H_J(t)$.
For two atoms in different Rydberg states, there is DD interaction between them, which scales as $1/R^3$ with $R$ being the distance between the two atoms~\cite{Browaeys2016,Browaeys2020}.
For sufficiently large lattice constant $a$, one can only consider the nearest-neighboring DD interaction.
The direct DD interaction between site $j$ and $j+1$ exchanges the Rydberg states in the two sites and reads $V_d[(|nS\rangle\langle n'P|)_j(|n'P\rangle\langle nS|)_{j+1}+h.c.]$ with $V_d=C_3/a^3$ and $C_3\sim n^4$ being the DD interaction coefficient~\cite{Browaeys2016}.
It simulates the tunneling term in our model.
To make the tunneling strength vary in time periodically, we propose a Raman coupling scheme, which contains two Raman lasers coupling the state $|nS\rangle$ to a low-lying state $|\alpha \rangle$, as described by
\begin{equation}
V_R(t)=\sum_j\{[\Omega_1\cos(\omega_{\mathrm{R}_1}t)+\Omega_2\cos(\omega_{\mathrm{R}_2}t)](|nS\rangle\langle \alpha|)_j+h.c.\}.
\end{equation}
Here $\Omega_{1,2}$ are the Rabi frequency and $\omega_{\mathrm{R}_1,\mathrm{R}_2}$ are the frequency of the two Raman lasers.
We suppose that the tilting strength $\tilde g$ is much larger than $V_d$ and the bare exchange process is suppressed due to this energy offset.
In the same spirit of Ref.~\cite{Yang2022}, we utilize the two-photon Raman process to compensate part of the energy offset with $\tilde g-(\omega_{\mathrm{R}_2}-\omega_{\mathrm{R}_1})=g$ and recover the exchange coupling.

There are two intermediate states $|n'P\rangle_j|\alpha\rangle_{j+1}$ and $|\alpha\rangle_j|n'P\rangle_{j+1}$ that help the coupling between $|nS\rangle_j|n'P\rangle_{j+1}$ and $|n'P\rangle_j|nS\rangle_{j+1}$, with the energy compensation happening at site $j+1$ and $j$ respectively, see Fig.~\ref{Fig_LDD}.
Labeling $|a\rangle_j=|nS\rangle_j|n'P\rangle_{j+1}$, $|b\rangle_j=|\alpha\rangle_j|n'P\rangle_{j+1}$, $|c\rangle_j=|n'P\rangle_j|\alpha\rangle_{j+1}$ and $|d\rangle_j=|n'P\rangle_j|nS\rangle_{j+1}$, the laser-assisted DD interaction in the four-level system under rotating wave approximation reads
\begin{widetext}
\begin{equation}
\hat H^{\mathrm{LDD}}(j,t)=
\begin{pmatrix}
E_a & \frac{\Omega_1}{2} e^{-i\omega_{\mathrm{R}_1}t}+\frac{\Omega_2}{2} e^{-i\omega_{\mathrm{R}_2}t} & 0 & V_d \\
\frac{\Omega_1^*}{2}e^{i\omega_{\mathrm{R}_1}t}+\frac{\Omega_2^*}{2}e^{i\omega_{\mathrm{R}_2}t} & E_b & 0 & 0 \\
0 & 0 & E_c & \frac{\Omega_1^*}{2}e^{i\omega_{\mathrm{R}1}t}+\frac{\Omega_2^*}{2}e^{i\omega_{\mathrm{R}2}t} \\
V_d & 0 & \frac{\Omega_1}{2} e^{-i\omega_{\mathrm{R}_1}t}+\frac{\Omega_2}{2} e^{-i\omega_{\mathrm{R}_2}t} & E_d
\end{pmatrix}
,
\end{equation}
where $E_d-E_a=\tilde g$, $E_a-E_b=E_d-E_c=\Delta$ and $\Delta$ is the energy difference between $|nS\rangle$ and $|\alpha\rangle$.
Transforming to the interaction picture, the Hamiltonian reads
\begin{equation}
\hat H^{\mathrm{LDD}}_{\mathrm{I}}(j,t)=
\begin{pmatrix}
0 & \frac{\Omega_1}{2} e^{i\delta_1 t}+\frac{\Omega_2}{2} e^{i\delta_2 t}  & 0 & V_de^{-i\tilde gt} \\
\frac{\Omega_1^*}{2} e^{-i\delta_1 t}+\frac{\Omega_2^*}{2} e^{-i\delta_2 t} & 0 & 0 & 0\\
0 & 0 & 0 & \frac{\Omega_1^*}{2}e^{-i\delta_1 t}+\frac{\Omega_2^*}{2}e^{-i\delta_2 t} \\
V_de^{i\tilde gt} & 0 & \frac{\Omega_1}{2} e^{i\delta_1 t}+\frac{\Omega_2}{2} e^{i\delta_2 t} & 0
\end{pmatrix}
,\nonumber\\
\end{equation}
with $\delta_1=\Delta-\omega_{\mathrm{R}_1}$ and $\delta_2=\Delta-\omega_{\mathrm{R}_2}$ being the single-photon detunings.
In the interaction picture, the evolution operator can be expanded in the Dyson series as
\begin{eqnarray}
\hat U_{\mathrm{I}}(j,t)&=&1-i\int_0^t\hat H^{\mathrm{LDD}}_{\mathrm{I}}(j,t_1)dt_1+(-i)^2\int_0^{t}\hat H^{\mathrm{LDD}}_{\mathrm{I}}(j,t_1)\int_0^{t_1}\hat H^{\mathrm{LDD}}_{\mathrm{I}}(j,t_2)dt_2dt_1\nonumber\\
&+&(-i)^3\int_0^t\hat H^{\mathrm{LDD}}_{\mathrm{I}}(j,t_1)\int_0^{t_1}\hat H^{\mathrm{LDD}}_{\mathrm{I}}(j,t_2)\int_0^{t_2}\hat H^{\mathrm{LDD}}_{\mathrm{I}}(j,t_3)dt_3dt_2dt_1+\cdots.
\end{eqnarray}
\end{widetext}

Suppose $\tilde g,|\delta_1|,|\delta_2|,|\tilde g\pm\delta_1|,|\tilde g\pm\delta_2|,|\delta_1-\delta_2|\gg V_d,|\Omega_1|,|\Omega_2|$ and $\tilde g\sim\omega_{\mathrm{R2}}-\omega_{R1}$.
The first- and second-order processes are non-resonant.
In the third-order processes, the couplings between $\{|a\rangle_j,|d\rangle_j\}$ and $\{|b\rangle_j,|c\rangle_j\}$ are non-resonant, and there are near-resonant couplings between $|a\rangle_j(|b\rangle_j)$ and $|d\rangle_j(|c\rangle_j)$.
Thus, the evolution operator in the interaction picture is block diagonalized.
According to the near-resonant processes in Fig.~\ref{Fig_LDD}, we can calculate the element matrix
\begin{equation}
U_{\mathrm{I}}^{(ad)}=\ _j\langle d|\hat U_{\mathrm{I}}(j,t)|a\rangle_j\nonumber\\
=J_{\mathrm{eff}}\frac{e^{-i(\omega_{21}-\tilde g)t}-1}{\omega_{21}-\tilde g},
\end{equation}
where $\omega_{21}=\omega_{\mathrm{R}_2}-\omega_{\mathrm{R}_1}$, and
\begin{equation}
J_{\mathrm{eff}}=\frac{V_d\Omega_1^*\Omega_2}{4}\left[\frac{1}{\delta_1\omega_{21}}+\frac{1}{\tilde g(\tilde g-\delta_1)}\right].
\end{equation}
The evolution operator in the bases $\{|a\rangle_j,|d\rangle_j\}$ up to third-order Dyson expansion then reads
\begin{equation}
\hat U_{\mathrm{I}}^{(ad)}(j,t)\approx1+\left[U_{\mathrm{I}}^{(ad)}(|d\rangle\langle a|)_j+h.c.\right].
\end{equation}
Transforming to the Schr\"{o}dinger picture, we can obtain the Hamiltonian
\begin{eqnarray}
&\hat H^{(ad)}&(j,t)=\hat H_0(j)+ie^{-i\hat H_0(j)t}\frac{\partial \hat U^{(ad)}_{\mathrm{I}}}{\partial t}{\hat U}^{(ad)\dagger}_{\mathrm{I}}e^{i\hat H_0(j)t}\nonumber\\
&=&\hat H_0(j)+J_{\mathrm{eff}}e^{-i(\omega_{21}-\tilde g+E_d-E_a)t}(|d\rangle\langle a|)_j+h.c.,\nonumber\\
&&
\end{eqnarray}
where $\hat H_0(j)=E_a(|a\rangle\langle a|)_j+E_d(|d\rangle\langle d|)_j$.
At last, we perform a rotating transformation and obtain the effective Hamiltonian
\begin{eqnarray}
\hat H^{(ad)}_{\mathrm{eff}}(j)&=&\hat S(t)\hat H^{(ad)}(j,t)\hatd S(t)
-i\hat S(t)
\frac{\partial \hatd S(t)}{\partial t} \nonumber\\
&=&-\frac{\tilde g-\omega_{21}}{2}(|a\rangle\langle a|)_j+\frac{\tilde g-\omega_{21}}{2}(|d\rangle\langle d|)_j\nonumber\\
&+&J_{\mathrm{eff}}(|d\rangle\langle a|)_j+h.c.,
\end{eqnarray}
where
\begin{equation}
\hat S(t)=\begin{pmatrix}
e^{i(E_a-\frac{\omega_{21}-\tilde g}{2})t} & 0 \\
0 & e^{i(E_d+\frac{\omega_{21}-\tilde g}{2})t}.
\end{pmatrix}
\end{equation}
If we set $\Omega_1=\Omega_2=\Omega$, and tune the Rabi frequency as $|\Omega|^2=|\Omega_0|^2|[1+u\cos(\omega t)]$, with the cosinoidal driving been realized experimentally~\cite{Omran2019}, the sum of $\hat H_{\mathrm{eff}}^{(ad)}(j)$ over $j$ simulates $\hat H_g+\hat H_J(t)$ with $J=V_d|\Omega_0|^2[\frac{1}{4\delta_1\omega_{21}}+\frac{1}{4\tilde g(\tilde g-\delta_1)}]$ and $g=\tilde g-\omega_{21}$.

The van der Waals interaction between Rydberg atoms can be used to simulate $\hat H_U$.
The atoms in the same Rydberg states can interact through the van der Waals interaction, which scales as $1/R^6$~\cite{Browaeys2016,Browaeys2020}.
If the lattice constant is sufficiently large, one can only consider the nearest-neighboring interaction.
The sum of the van der Walls interactions of the two kinds of Rydberg states read
\begin{eqnarray}
\hat H_{\mathrm{vdW}}&=&\frac{C_6^{(2)}+C_6^{(1)}}{a^6}\sum_{j=0}^{L-2}\left(|n'P\rangle\langle n'P|\right)_j\left(|n'P\rangle\langle n'P|\right)_{j+1}\nonumber\\
&+&\frac{C_6^{(1)}}{a^6}\left[\left(|n'P\rangle\langle n'P|\right)_0+\left(|n'P\rangle\langle n'P|\right)_{L-1}\right],
\end{eqnarray}
where $C_6^{(1)}\sim n^{11}$ and $C_6^{(2)}\sim n'^{11}$ are the van der Waals coefficients of $|nS\rangle$ and $|n'P\rangle$~\cite{Browaeys2016,Browaeys2020}, and a constant energy $-LC_6^{(1)}/a^6$ has been dropped.
The defect on the boundary sites can be compensated by light shift.
Thus, the van der Walls interaction simulates $\hat H_U$ with $U=[C_6^{(2)}+C_6^{(1)}]/a^6$.

\section{The odd-even effect of particle number on the largest Krylov subspace when $\omega=\omega_2$}\label{AppA}
The spatial-reflection operator reflects the particles on a Fock state $|n_0n_1\cdots n_{L-2} n_{L-1}\rangle$, such that $\hat R|\vec n\rangle=|n_{L-1}n_{L-2}\cdots n_1n_0\rangle\equiv|\underline{\vec n}\rangle$.
It is easy to see that
\begin{eqnarray}
\hat{\mathcal P}^{(2g)}_{L-j-1,L-j+2}\hatd c_{L-j+1}\hat c_{L-j}&=&\hat R\hat{\mathcal P}^{(0)}_{j-1,j+2}\hatd c_j\hat c_{j+1}\hat R^{-1}, \nonumber\\
\hat{\mathcal P}^{(2g)}_{L-j-1,L-j+2}\hatd c_{L-j}\hat c_{L-j+1}&=&\hat R\hat{\mathcal P}^{(0)}_{j-1,j+2}\hatd c_{j+1}\hat c_j\hat R^{-1}.\nonumber\\
&&
\end{eqnarray}
Thus, corresponding to a nonzero $\langle \vec l|\hat H^{\mathrm{eff}}_{\omega_2}|\vec n\rangle$, the matrix element $\langle \vec{\underline l}|\hat H^{\mathrm{eff}}_{\omega_2}|\vec{\underline n}\rangle$ is also nonzero.
It is obvious that if a Krylov subspace contains a pair of mutually reflected states $\{|\vec n\rangle,|\underline{\vec n}\rangle\}$ (including the case when $|\vec n\rangle$ is a reflection invariant state with $|\underline{\vec n}\rangle=|\vec n\rangle$), then all the Fock states in this Krylov subspace can be paired by the spatial reflection transformation.
Otherwise, the Krylov subspace does not contain any pair of mutually reflected Fock states.
Under the spatial reflection, the Krylov subspace of the former kind is invariant, in the sense that the set of bases in the Krylov subspace does not change.
The Krylov subspace of the the latter kind will be transformed to its reflection partner.

Depending on the parity of the total particle number $N$, there will exist reflection invariant Krylov subspace or not.
The expectation values of $\hat E$ in a Fock state $|\vec n\rangle$ and its reflected state read $E_{\vec n}=\langle \vec n|\hat E|\vec n\rangle=-\sum_jjn_j+\sum_jn_jn_{j+1}$ and $E_{\vec{\underline n}}=E_{\vec n}-2\sum_jjn_{L+1-j}+N(L+1)$, respectively.
If $N$ is odd, the parity of $\hat E$ in the two states are opposite to each other, thus $|\vec n\rangle$ and $|\vec{\underline n}\rangle$ belong to different parity symmetry spaces.
In this way, the Krylov subspaces in the even- and odd-parity symmetry spaces are in one-to-one correspondence through the spatial-reflection transformation, and there is no reflection invariant Krylov subspace.
If $N$ is even, the parity of $\hat E$ in a pair of mutually reflected Fock states are the same, and there exists reflection invariant Krylov subspaces.
One such Krylov subspace is the one that contains the charge density wave state $|\mathrm{CDW}1\rangle=|0101\cdots01\rangle$ and its reflected partner $|\mathrm{CDW}2\rangle=|1010\cdots10\rangle$.
Viewing four sites as a cell, both the sequences $0101$ in $|\mathrm{CDW}1\rangle$ and $1010$ in $|\mathrm{CDW}2\rangle$ can be changed to $0110$ under the action of $\hat H^{\mathrm{eff}}_{\omega_2}$.
Thus, under $N/2$ times of action of $\hat H^{\mathrm{eff}}_{\omega_2}$, both $|\mathrm{CDW}1\rangle$ and $|\mathrm{CDW2}\rangle$ can be couple to the same state
$|01100110\cdots0110\rangle$, which is reflection invariant.
So, $|\mathrm{CDW}1\rangle$ and $|\mathrm{CDW2}\rangle$ are in the same Krylov subspace and this subspace is reflection invariant.

Numerically, we find that the largest Krylov subspace at all system sizes contains the charge density wave states.
Thus, when $N$ is odd, there are two largest Krylov subspace in the half-filling sector, which lie in the even- and odd-parity symmetry space of $\hat E$ respectively.
When $N$ is even, there is only one largest Krylov subspace, which lies in the even-parity symmetry space and is reflection invariant [note that $E_{\mathrm{CDW}1}=-N^2$ and $E_{\mathrm{CDW}2}=-N(N-1)$].
This odd-even effect leads to different fitting functions for quadruple and non-quadruple lattice sites.


\begin{thebibliography}{58}%
\makeatletter
\providecommand \@ifxundefined [1]{%
 \@ifx{#1\undefined}
}%
\providecommand \@ifnum [1]{%
 \ifnum #1\expandafter \@firstoftwo
 \else \expandafter \@secondoftwo
 \fi
}%
\providecommand \@ifx [1]{%
 \ifx #1\expandafter \@firstoftwo
 \else \expandafter \@secondoftwo
 \fi
}%
\providecommand \natexlab [1]{#1}%
\providecommand \enquote  [1]{``#1''}%
\providecommand \bibnamefont  [1]{#1}%
\providecommand \bibfnamefont [1]{#1}%
\providecommand \citenamefont [1]{#1}%
\providecommand \href@noop [0]{\@secondoftwo}%
\providecommand \href [0]{\begingroup \@sanitize@url \@href}%
\providecommand \@href[1]{\@@startlink{#1}\@@href}%
\providecommand \@@href[1]{\endgroup#1\@@endlink}%
\providecommand \@sanitize@url [0]{\catcode `\\12\catcode `\$12\catcode
  `\&12\catcode `\#12\catcode `\^12\catcode `\_12\catcode `\%12\relax}%
\providecommand \@@startlink[1]{}%
\providecommand \@@endlink[0]{}%
\providecommand \url  [0]{\begingroup\@sanitize@url \@url }%
\providecommand \@url [1]{\endgroup\@href {#1}{\urlprefix }}%
\providecommand \urlprefix  [0]{URL }%
\providecommand \Eprint [0]{\href }%
\providecommand \doibase [0]{http://dx.doi.org/}%
\providecommand \selectlanguage [0]{\@gobble}%
\providecommand \bibinfo  [0]{\@secondoftwo}%
\providecommand \bibfield  [0]{\@secondoftwo}%
\providecommand \translation [1]{[#1]}%
\providecommand \BibitemOpen [0]{}%
\providecommand \bibitemStop [0]{}%
\providecommand \bibitemNoStop [0]{.\EOS\space}%
\providecommand \EOS [0]{\spacefactor3000\relax}%
\providecommand \BibitemShut  [1]{\csname bibitem#1\endcsname}%
\let\auto@bib@innerbib\@empty
\bibitem [{\citenamefont {Deutsch}(1991)}]{Deutsch1991}%
  \BibitemOpen
  \bibfield  {author} {\bibinfo {author} {\bibfnamefont {J.~M.}\ \bibnamefont
  {Deutsch}},\ }\href {\doibase 10.1103/PhysRevA.43.2046} {\bibfield  {journal}
  {\bibinfo  {journal} {Phys. Rev. A}\ }\textbf {\bibinfo {volume} {43}},\
  \bibinfo {pages} {2046} (\bibinfo {year} {1991})}\BibitemShut {NoStop}%
\bibitem [{\citenamefont {Srednicki}(1994)}]{Srednicki1994}%
  \BibitemOpen
  \bibfield  {author} {\bibinfo {author} {\bibfnamefont {M.}~\bibnamefont
  {Srednicki}},\ }\href {\doibase 10.1103/PhysRevE.50.888} {\bibfield
  {journal} {\bibinfo  {journal} {Phys. Rev. E}\ }\textbf {\bibinfo {volume}
  {50}},\ \bibinfo {pages} {888} (\bibinfo {year} {1994})}\BibitemShut
  {NoStop}%
\bibitem [{\citenamefont {Srednicki}(1999)}]{Srednicki1999}%
  \BibitemOpen
  \bibfield  {author} {\bibinfo {author} {\bibfnamefont {M.}~\bibnamefont
  {Srednicki}},\ }\href {\doibase 10.1088/0305-4470/32/7/007} {\bibfield
  {journal} {\bibinfo  {journal} {J. Phys. A}\ }\textbf {\bibinfo {volume}
  {32}},\ \bibinfo {pages} {1163} (\bibinfo {year} {1999})}\BibitemShut
  {NoStop}%
\bibitem [{\citenamefont {Rigol}\ and\ \citenamefont
  {Dunjko}(2008)}]{Rigol2008}%
  \BibitemOpen
  \bibfield  {author} {\bibinfo {author} {\bibfnamefont {M.}~\bibnamefont
  {Rigol}}\ and\ \bibinfo {author} {\bibfnamefont {M.}~\bibnamefont {Dunjko},
  \bibfnamefont {V.~andn~Olshanii}},\ }\href {\doibase 10.1038/nature06838}
  {\bibfield  {journal} {\bibinfo  {journal} {Nature}\ }\textbf {\bibinfo
  {volume} {452}},\ \bibinfo {pages} {854} (\bibinfo {year}
  {2008})}\BibitemShut {NoStop}%
\bibitem [{\citenamefont {Abanin}\ \emph {et~al.}(2016)\citenamefont {Abanin},
  \citenamefont {Roeck},\ and\ \citenamefont {Huveneers}}]{Abanin2016}%
  \BibitemOpen
  \bibfield  {author} {\bibinfo {author} {\bibfnamefont {D.~A.}\ \bibnamefont
  {Abanin}}, \bibinfo {author} {\bibfnamefont {W.~D.}\ \bibnamefont {Roeck}}, \
  and\ \bibinfo {author} {\bibfnamefont {F.}~\bibnamefont {Huveneers}},\ }\href
  {\doibase https://doi.org/10.1016/j.aop.2016.03.010} {\bibfield  {journal}
  {\bibinfo  {journal} {Ann. Phys.}\ }\textbf {\bibinfo {volume} {372}},\
  \bibinfo {pages} {1} (\bibinfo {year} {2016})}\BibitemShut {NoStop}%
\bibitem [{\citenamefont {Nandkishore}\ and\ \citenamefont
  {Huse}(2015)}]{Nandkishore2015}%
  \BibitemOpen
  \bibfield  {author} {\bibinfo {author} {\bibfnamefont {R.}~\bibnamefont
  {Nandkishore}}\ and\ \bibinfo {author} {\bibfnamefont {D.~A.}\ \bibnamefont
  {Huse}},\ }\href {\doibase 10.1146/annurev-conmatphys-031214-014726}
  {\bibfield  {journal} {\bibinfo  {journal} {Annu. Rev. Condens. Matter.
  Phys.}\ }\textbf {\bibinfo {volume} {6}},\ \bibinfo {pages} {15} (\bibinfo
  {year} {2015})}\BibitemShut {NoStop}%
\bibitem [{\citenamefont {Abanin}\ \emph {et~al.}(2019)\citenamefont {Abanin},
  \citenamefont {Altman}, \citenamefont {Bloch},\ and\ \citenamefont
  {Serbyn}}]{Abanin2019}%
  \BibitemOpen
  \bibfield  {author} {\bibinfo {author} {\bibfnamefont {D.~A.}\ \bibnamefont
  {Abanin}}, \bibinfo {author} {\bibfnamefont {E.}~\bibnamefont {Altman}},
  \bibinfo {author} {\bibfnamefont {I.}~\bibnamefont {Bloch}}, \ and\ \bibinfo
  {author} {\bibfnamefont {M.}~\bibnamefont {Serbyn}},\ }\href {\doibase
  10.1103/RevModPhys.91.021001} {\bibfield  {journal} {\bibinfo  {journal}
  {Rev. Mod. Phys.}\ }\textbf {\bibinfo {volume} {91}},\ \bibinfo {pages}
  {021001} (\bibinfo {year} {2019})}\BibitemShut {NoStop}%
\bibitem [{\citenamefont {Moudgalya}\ \emph {et~al.}(2022)\citenamefont
  {Moudgalya}, \citenamefont {Bernevig},\ and\ \citenamefont
  {Regnault}}]{Moudgalya2022}%
  \BibitemOpen
  \bibfield  {author} {\bibinfo {author} {\bibfnamefont {S.}~\bibnamefont
  {Moudgalya}}, \bibinfo {author} {\bibfnamefont {B.~A.}\ \bibnamefont
  {Bernevig}}, \ and\ \bibinfo {author} {\bibfnamefont {N.}~\bibnamefont
  {Regnault}},\ }\href {\doibase 10.1088/1361-6633/ac73a0} {\bibfield
  {journal} {\bibinfo  {journal} {Rep. Prog. Phys.}\ }\textbf {\bibinfo
  {volume} {85}},\ \bibinfo {pages} {086501} (\bibinfo {year}
  {2022})}\BibitemShut {NoStop}%
\bibitem [{\citenamefont {Kinoshita}\ \emph {et~al.}(2006)\citenamefont
  {Kinoshita}, \citenamefont {Wenger},\ and\ \citenamefont
  {Weiss}}]{Kinoshita2006}%
  \BibitemOpen
  \bibfield  {author} {\bibinfo {author} {\bibfnamefont {T.}~\bibnamefont
  {Kinoshita}}, \bibinfo {author} {\bibfnamefont {T.}~\bibnamefont {Wenger}}, \
  and\ \bibinfo {author} {\bibfnamefont {D.~S.}\ \bibnamefont {Weiss}},\ }\href
  {\doibase 10.1038/nature04693} {\bibfield  {journal} {\bibinfo  {journal}
  {Nature}\ }\textbf {\bibinfo {volume} {440}},\ \bibinfo {pages} {900}
  (\bibinfo {year} {2006})}\BibitemShut {NoStop}%
\bibitem [{\citenamefont {Rigol}\ \emph {et~al.}(2007)\citenamefont {Rigol},
  \citenamefont {Dunjko}, \citenamefont {Yurovsky},\ and\ \citenamefont
  {Olshanii}}]{Rigol2007}%
  \BibitemOpen
  \bibfield  {author} {\bibinfo {author} {\bibfnamefont {M.}~\bibnamefont
  {Rigol}}, \bibinfo {author} {\bibfnamefont {V.}~\bibnamefont {Dunjko}},
  \bibinfo {author} {\bibfnamefont {V.}~\bibnamefont {Yurovsky}}, \ and\
  \bibinfo {author} {\bibfnamefont {M.}~\bibnamefont {Olshanii}},\ }\href
  {\doibase 10.1103/PhysRevLett.98.050405} {\bibfield  {journal} {\bibinfo
  {journal} {Phys. Rev. Lett.}\ }\textbf {\bibinfo {volume} {98}},\ \bibinfo
  {pages} {050405} (\bibinfo {year} {2007})}\BibitemShut {NoStop}%
\bibitem [{\citenamefont {Polkovnikov}\ \emph {et~al.}(2011)\citenamefont
  {Polkovnikov}, \citenamefont {Sengupta}, \citenamefont {Silva},\ and\
  \citenamefont {Vengalattore}}]{Polkovnikov2011}%
  \BibitemOpen
  \bibfield  {author} {\bibinfo {author} {\bibfnamefont {A.}~\bibnamefont
  {Polkovnikov}}, \bibinfo {author} {\bibfnamefont {K.}~\bibnamefont
  {Sengupta}}, \bibinfo {author} {\bibfnamefont {A.}~\bibnamefont {Silva}}, \
  and\ \bibinfo {author} {\bibfnamefont {M.}~\bibnamefont {Vengalattore}},\
  }\href {\doibase 10.1103/RevModPhys.83.863} {\bibfield  {journal} {\bibinfo
  {journal} {Rev. Mod. Phys.}\ }\textbf {\bibinfo {volume} {83}},\ \bibinfo
  {pages} {863} (\bibinfo {year} {2011})}\BibitemShut {NoStop}%
\bibitem [{\citenamefont {Serbyn}\ \emph {et~al.}(2013)\citenamefont {Serbyn},
  \citenamefont {Papi\ifmmode~\acute{c}\else \'{c}\fi{}},\ and\ \citenamefont
  {Abanin}}]{Serbyn2013}%
  \BibitemOpen
  \bibfield  {author} {\bibinfo {author} {\bibfnamefont {M.}~\bibnamefont
  {Serbyn}}, \bibinfo {author} {\bibfnamefont {Z.}~\bibnamefont
  {Papi\ifmmode~\acute{c}\else \'{c}\fi{}}}, \ and\ \bibinfo {author}
  {\bibfnamefont {D.~A.}\ \bibnamefont {Abanin}},\ }\href {\doibase
  10.1103/PhysRevLett.111.127201} {\bibfield  {journal} {\bibinfo  {journal}
  {Phys. Rev. Lett.}\ }\textbf {\bibinfo {volume} {111}},\ \bibinfo {pages}
  {127201} (\bibinfo {year} {2013})}\BibitemShut {NoStop}%
\bibitem [{\citenamefont {Huse}\ \emph {et~al.}(2014)\citenamefont {Huse},
  \citenamefont {Nandkishore},\ and\ \citenamefont {Oganesyan}}]{Huse2014}%
  \BibitemOpen
  \bibfield  {author} {\bibinfo {author} {\bibfnamefont {D.~A.}\ \bibnamefont
  {Huse}}, \bibinfo {author} {\bibfnamefont {R.}~\bibnamefont {Nandkishore}}, \
  and\ \bibinfo {author} {\bibfnamefont {V.}~\bibnamefont {Oganesyan}},\ }\href
  {\doibase 10.1103/PhysRevB.90.174202} {\bibfield  {journal} {\bibinfo
  {journal} {Phys. Rev. B}\ }\textbf {\bibinfo {volume} {90}},\ \bibinfo
  {pages} {174202} (\bibinfo {year} {2014})}\BibitemShut {NoStop}%
\bibitem [{\citenamefont {Schulz}\ \emph {et~al.}(2019)\citenamefont {Schulz},
  \citenamefont {Hooley}, \citenamefont {Moessner},\ and\ \citenamefont
  {Pollmann}}]{Schulz2019}%
  \BibitemOpen
  \bibfield  {author} {\bibinfo {author} {\bibfnamefont {M.}~\bibnamefont
  {Schulz}}, \bibinfo {author} {\bibfnamefont {C.~A.}\ \bibnamefont {Hooley}},
  \bibinfo {author} {\bibfnamefont {R.}~\bibnamefont {Moessner}}, \ and\
  \bibinfo {author} {\bibfnamefont {F.}~\bibnamefont {Pollmann}},\ }\href
  {\doibase 10.1103/PhysRevLett.122.040606} {\bibfield  {journal} {\bibinfo
  {journal} {Phys. Rev. Lett.}\ }\textbf {\bibinfo {volume} {122}},\ \bibinfo
  {pages} {040606} (\bibinfo {year} {2019})}\BibitemShut {NoStop}%
\bibitem [{\citenamefont {van Nieuwenburg}\ \emph {et~al.}(2019)\citenamefont
  {van Nieuwenburg}, \citenamefont {Baum},\ and\ \citenamefont
  {Refael}}]{Nieuwenburg2019}%
  \BibitemOpen
  \bibfield  {author} {\bibinfo {author} {\bibfnamefont {E.}~\bibnamefont {van
  Nieuwenburg}}, \bibinfo {author} {\bibfnamefont {Y.}~\bibnamefont {Baum}}, \
  and\ \bibinfo {author} {\bibfnamefont {G.}~\bibnamefont {Refael}},\ }\href
  {\doibase 10.1073/pnas.1819316116} {\bibfield  {journal} {\bibinfo  {journal}
  {Proc. Natl. Acad. Sci. U.S.A.}\ }\textbf {\bibinfo {volume} {116}},\
  \bibinfo {pages} {9269} (\bibinfo {year} {2019})}\BibitemShut {NoStop}%
\bibitem [{\citenamefont {Taylor}\ \emph {et~al.}(2020)\citenamefont {Taylor},
  \citenamefont {Schulz}, \citenamefont {Pollmann},\ and\ \citenamefont
  {Moessner}}]{Taylor2020}%
  \BibitemOpen
  \bibfield  {author} {\bibinfo {author} {\bibfnamefont {S.~R.}\ \bibnamefont
  {Taylor}}, \bibinfo {author} {\bibfnamefont {M.}~\bibnamefont {Schulz}},
  \bibinfo {author} {\bibfnamefont {F.}~\bibnamefont {Pollmann}}, \ and\
  \bibinfo {author} {\bibfnamefont {R.}~\bibnamefont {Moessner}},\ }\href
  {\doibase 10.1103/PhysRevB.102.054206} {\bibfield  {journal} {\bibinfo
  {journal} {Phys. Rev. B}\ }\textbf {\bibinfo {volume} {102}},\ \bibinfo
  {pages} {054206} (\bibinfo {year} {2020})}\BibitemShut {NoStop}%
\bibitem [{\citenamefont {Yao}\ and\ \citenamefont
  {Zakrzewski}(2020)}]{Yao2020}%
  \BibitemOpen
  \bibfield  {author} {\bibinfo {author} {\bibfnamefont {R.}~\bibnamefont
  {Yao}}\ and\ \bibinfo {author} {\bibfnamefont {J.}~\bibnamefont
  {Zakrzewski}},\ }\href {\doibase 10.1103/PhysRevB.102.104203} {\bibfield
  {journal} {\bibinfo  {journal} {Phys. Rev. B}\ }\textbf {\bibinfo {volume}
  {102}},\ \bibinfo {pages} {104203} (\bibinfo {year} {2020})}\BibitemShut
  {NoStop}%
\bibitem [{\citenamefont {Chanda}\ \emph {et~al.}(2020)\citenamefont {Chanda},
  \citenamefont {Yao},\ and\ \citenamefont {Zakrzewski}}]{Chanda2020}%
  \BibitemOpen
  \bibfield  {author} {\bibinfo {author} {\bibfnamefont {T.}~\bibnamefont
  {Chanda}}, \bibinfo {author} {\bibfnamefont {R.}~\bibnamefont {Yao}}, \ and\
  \bibinfo {author} {\bibfnamefont {J.}~\bibnamefont {Zakrzewski}},\ }\href
  {\doibase 10.1103/PhysRevResearch.2.032039} {\bibfield  {journal} {\bibinfo
  {journal} {Phys. Rev. Research}\ }\textbf {\bibinfo {volume} {2}},\ \bibinfo
  {pages} {032039} (\bibinfo {year} {2020})}\BibitemShut {NoStop}%
\bibitem [{\citenamefont {Zhang}\ \emph {et~al.}(2021)\citenamefont {Zhang},
  \citenamefont {Ke}, \citenamefont {Liu},\ and\ \citenamefont
  {Lee}}]{Zhang2021}%
  \BibitemOpen
  \bibfield  {author} {\bibinfo {author} {\bibfnamefont {L.}~\bibnamefont
  {Zhang}}, \bibinfo {author} {\bibfnamefont {Y.}~\bibnamefont {Ke}}, \bibinfo
  {author} {\bibfnamefont {W.}~\bibnamefont {Liu}}, \ and\ \bibinfo {author}
  {\bibfnamefont {C.}~\bibnamefont {Lee}},\ }\href {\doibase
  10.1103/PhysRevA.103.023323} {\bibfield  {journal} {\bibinfo  {journal}
  {Phys. Rev. A}\ }\textbf {\bibinfo {volume} {103}},\ \bibinfo {pages}
  {023323} (\bibinfo {year} {2021})}\BibitemShut {NoStop}%
\bibitem [{\citenamefont {Yao}\ \emph {et~al.}(2021)\citenamefont {Yao},
  \citenamefont {Chanda},\ and\ \citenamefont {Zakrzewski}}]{Yao2021}%
  \BibitemOpen
  \bibfield  {author} {\bibinfo {author} {\bibfnamefont {R.}~\bibnamefont
  {Yao}}, \bibinfo {author} {\bibfnamefont {T.}~\bibnamefont {Chanda}}, \ and\
  \bibinfo {author} {\bibfnamefont {J.}~\bibnamefont {Zakrzewski}},\ }\href
  {\doibase 10.1103/PhysRevB.104.014201} {\bibfield  {journal} {\bibinfo
  {journal} {Phys. Rev. B}\ }\textbf {\bibinfo {volume} {104}},\ \bibinfo
  {pages} {014201} (\bibinfo {year} {2021})}\BibitemShut {NoStop}%
\bibitem [{\citenamefont {Wang}\ \emph {et~al.}(2021)\citenamefont {Wang},
  \citenamefont {Sun},\ and\ \citenamefont {Fan}}]{Wang2021}%
  \BibitemOpen
  \bibfield  {author} {\bibinfo {author} {\bibfnamefont {Y.-Y.}\ \bibnamefont
  {Wang}}, \bibinfo {author} {\bibfnamefont {Z.-H.}\ \bibnamefont {Sun}}, \
  and\ \bibinfo {author} {\bibfnamefont {H.}~\bibnamefont {Fan}},\ }\href
  {\doibase 10.1103/PhysRevB.104.205122} {\bibfield  {journal} {\bibinfo
  {journal} {Phys. Rev. B}\ }\textbf {\bibinfo {volume} {104}},\ \bibinfo
  {pages} {205122} (\bibinfo {year} {2021})}\BibitemShut {NoStop}%
\bibitem [{\citenamefont {Guo}\ \emph {et~al.}(2021)\citenamefont {Guo},
  \citenamefont {Cheng}, \citenamefont {Li}, \citenamefont {Xu}, \citenamefont
  {Zhang}, \citenamefont {Wang}, \citenamefont {Song}, \citenamefont {Liu},
  \citenamefont {Ren}, \citenamefont {Dong}, \citenamefont {Mondaini},\ and\
  \citenamefont {Wang}}]{Guo2021}%
  \BibitemOpen
  \bibfield  {author} {\bibinfo {author} {\bibfnamefont {Q.}~\bibnamefont
  {Guo}}, \bibinfo {author} {\bibfnamefont {C.}~\bibnamefont {Cheng}}, \bibinfo
  {author} {\bibfnamefont {H.}~\bibnamefont {Li}}, \bibinfo {author}
  {\bibfnamefont {S.}~\bibnamefont {Xu}}, \bibinfo {author} {\bibfnamefont
  {P.}~\bibnamefont {Zhang}}, \bibinfo {author} {\bibfnamefont
  {Z.}~\bibnamefont {Wang}}, \bibinfo {author} {\bibfnamefont {C.}~\bibnamefont
  {Song}}, \bibinfo {author} {\bibfnamefont {W.}~\bibnamefont {Liu}}, \bibinfo
  {author} {\bibfnamefont {W.}~\bibnamefont {Ren}}, \bibinfo {author}
  {\bibfnamefont {H.}~\bibnamefont {Dong}}, \bibinfo {author} {\bibfnamefont
  {R.}~\bibnamefont {Mondaini}}, \ and\ \bibinfo {author} {\bibfnamefont
  {H.}~\bibnamefont {Wang}},\ }\href {\doibase 10.1103/PhysRevLett.127.240502}
  {\bibfield  {journal} {\bibinfo  {journal} {Phys. Rev. Lett.}\ }\textbf
  {\bibinfo {volume} {127}},\ \bibinfo {pages} {240502} (\bibinfo {year}
  {2021})}\BibitemShut {NoStop}%
\bibitem [{\citenamefont {Morong}\ \emph {et~al.}(2021)\citenamefont {Morong},
  \citenamefont {Liu}, \citenamefont {Becher}, \citenamefont {Collins},
  \citenamefont {Feng}, \citenamefont {Kyprianidis}, \citenamefont {Pagano},
  \citenamefont {You}, \citenamefont {Gorshkov},\ and\ \citenamefont
  {Monroe}}]{Morong2021}%
  \BibitemOpen
  \bibfield  {author} {\bibinfo {author} {\bibfnamefont {W.}~\bibnamefont
  {Morong}}, \bibinfo {author} {\bibfnamefont {F.}~\bibnamefont {Liu}},
  \bibinfo {author} {\bibfnamefont {P.}~\bibnamefont {Becher}}, \bibinfo
  {author} {\bibfnamefont {K.~S.}\ \bibnamefont {Collins}}, \bibinfo {author}
  {\bibfnamefont {L.}~\bibnamefont {Feng}}, \bibinfo {author} {\bibfnamefont
  {A.}~\bibnamefont {Kyprianidis}}, \bibinfo {author} {\bibfnamefont
  {G.}~\bibnamefont {Pagano}}, \bibinfo {author} {\bibfnamefont
  {T.}~\bibnamefont {You}}, \bibinfo {author} {\bibfnamefont {A.~V.}\
  \bibnamefont {Gorshkov}}, \ and\ \bibinfo {author} {\bibfnamefont
  {C.}~\bibnamefont {Monroe}},\ }\href {\doibase
  https://doi.org/10.1038/s41586-021-03988-0} {\bibfield  {journal} {\bibinfo
  {journal} {Nature}\ }\textbf {\bibinfo {volume} {599}},\ \bibinfo {pages}
  {393} (\bibinfo {year} {2021})}\BibitemShut {NoStop}%
\bibitem [{\citenamefont {Turner}\ \emph
  {et~al.}(2018{\natexlab{a}})\citenamefont {Turner}, \citenamefont
  {Michailidis}, \citenamefont {Abanin}, \citenamefont {Serbyn},\ and\
  \citenamefont {Papi\ifmmode~\acute{c}\else \'{c}\fi{}}}]{Turner2018}%
  \BibitemOpen
  \bibfield  {author} {\bibinfo {author} {\bibfnamefont {C.~J.}\ \bibnamefont
  {Turner}}, \bibinfo {author} {\bibfnamefont {A.~A.}\ \bibnamefont
  {Michailidis}}, \bibinfo {author} {\bibfnamefont {D.~A.}\ \bibnamefont
  {Abanin}}, \bibinfo {author} {\bibfnamefont {M.}~\bibnamefont {Serbyn}}, \
  and\ \bibinfo {author} {\bibfnamefont {Z.}~\bibnamefont
  {Papi\ifmmode~\acute{c}\else \'{c}\fi{}}},\ }\href {\doibase
  10.1038/s41567-018-0137-5} {\bibfield  {journal} {\bibinfo  {journal} {Nat.
  Phys.}\ }\textbf {\bibinfo {volume} {14}},\ \bibinfo {pages} {745} (\bibinfo
  {year} {2018}{\natexlab{a}})}\BibitemShut {NoStop}%
\bibitem [{\citenamefont {Turner}\ \emph
  {et~al.}(2018{\natexlab{b}})\citenamefont {Turner}, \citenamefont
  {Michailidis}, \citenamefont {Abanin}, \citenamefont {Serbyn},\ and\
  \citenamefont {Papi\ifmmode~\acute{c}\else \'{c}\fi{}}}]{Turner2018_2}%
  \BibitemOpen
  \bibfield  {author} {\bibinfo {author} {\bibfnamefont {C.~J.}\ \bibnamefont
  {Turner}}, \bibinfo {author} {\bibfnamefont {A.~A.}\ \bibnamefont
  {Michailidis}}, \bibinfo {author} {\bibfnamefont {D.~A.}\ \bibnamefont
  {Abanin}}, \bibinfo {author} {\bibfnamefont {M.}~\bibnamefont {Serbyn}}, \
  and\ \bibinfo {author} {\bibfnamefont {Z.}~\bibnamefont
  {Papi\ifmmode~\acute{c}\else \'{c}\fi{}}},\ }\href {\doibase
  10.1103/PhysRevB.98.155134} {\bibfield  {journal} {\bibinfo  {journal} {Phys.
  Rev. B}\ }\textbf {\bibinfo {volume} {98}},\ \bibinfo {pages} {155134}
  (\bibinfo {year} {2018}{\natexlab{b}})}\BibitemShut {NoStop}%
\bibitem [{\citenamefont {Ho}\ \emph {et~al.}(2019)\citenamefont {Ho},
  \citenamefont {Choi}, \citenamefont {Pichler},\ and\ \citenamefont
  {Lukin}}]{Ho2019}%
  \BibitemOpen
  \bibfield  {author} {\bibinfo {author} {\bibfnamefont {W.}~\bibnamefont
  {Ho}}, \bibinfo {author} {\bibfnamefont {S.}~\bibnamefont {Choi}}, \bibinfo
  {author} {\bibfnamefont {H.}~\bibnamefont {Pichler}}, \ and\ \bibinfo
  {author} {\bibfnamefont {M.~D.}\ \bibnamefont {Lukin}},\ }\href {\doibase
  10.1103/PhysRevLett.122.040603} {\bibfield  {journal} {\bibinfo  {journal}
  {Phys. Rev. Lett.}\ }\textbf {\bibinfo {volume} {122}},\ \bibinfo {pages}
  {040603} (\bibinfo {year} {2019})}\BibitemShut {NoStop}%
\bibitem [{\citenamefont {Serbyn}\ \emph {et~al.}(2021)\citenamefont {Serbyn},
  \citenamefont {Abanin},\ and\ \citenamefont {Papi\ifmmode~\acute{c}\else
  \'{c}\fi{}}}]{Serbyn2021}%
  \BibitemOpen
  \bibfield  {author} {\bibinfo {author} {\bibfnamefont {M.}~\bibnamefont
  {Serbyn}}, \bibinfo {author} {\bibfnamefont {D.~A.}\ \bibnamefont {Abanin}},
  \ and\ \bibinfo {author} {\bibfnamefont {Z.}~\bibnamefont
  {Papi\ifmmode~\acute{c}\else \'{c}\fi{}}},\ }\href {\doibase
  https://doi.org/10.1038/s41567-021-01230-2} {\bibfield  {journal} {\bibinfo
  {journal} {Nat. Phys.}\ }\textbf {\bibinfo {volume} {17}},\ \bibinfo {pages}
  {675} (\bibinfo {year} {2021})}\BibitemShut {NoStop}%
\bibitem [{\citenamefont {Chandran}\ \emph {et~al.}(2023)\citenamefont
  {Chandran}, \citenamefont {Iadecola}, \citenamefont {Khemani},\ and\
  \citenamefont {Moessner}}]{Chandran2023}%
  \BibitemOpen
  \bibfield  {author} {\bibinfo {author} {\bibfnamefont {A.}~\bibnamefont
  {Chandran}}, \bibinfo {author} {\bibfnamefont {T.}~\bibnamefont {Iadecola}},
  \bibinfo {author} {\bibfnamefont {V.}~\bibnamefont {Khemani}}, \ and\
  \bibinfo {author} {\bibfnamefont {R.}~\bibnamefont {Moessner}},\ }\href
  {\doibase 10.1146/annurev-conmatphys-031620-101617} {\bibfield  {journal}
  {\bibinfo  {journal} {Annu. Rev. Condens. Matter. Phys.}\ }\textbf {\bibinfo
  {volume} {14}},\ \bibinfo {pages} {443} (\bibinfo {year} {2023})}\BibitemShut
  {NoStop}%
\bibitem [{\citenamefont {Pai}\ \emph {et~al.}(2019)\citenamefont {Pai},
  \citenamefont {Pretko},\ and\ \citenamefont {Nandkishore}}]{Pai2019}%
  \BibitemOpen
  \bibfield  {author} {\bibinfo {author} {\bibfnamefont {S.}~\bibnamefont
  {Pai}}, \bibinfo {author} {\bibfnamefont {M.}~\bibnamefont {Pretko}}, \ and\
  \bibinfo {author} {\bibfnamefont {R.~M.}\ \bibnamefont {Nandkishore}},\
  }\href {\doibase 10.1103/PhysRevX.9.021003} {\bibfield  {journal} {\bibinfo
  {journal} {Phys. Rev. X}\ }\textbf {\bibinfo {volume} {9}},\ \bibinfo {pages}
  {021003} (\bibinfo {year} {2019})}\BibitemShut {NoStop}%
\bibitem [{\citenamefont {Sala}\ \emph {et~al.}(2020)\citenamefont {Sala},
  \citenamefont {Rakovszky}, \citenamefont {Verresen}, \citenamefont {Knap},\
  and\ \citenamefont {Pollmann}}]{Sala2020}%
  \BibitemOpen
  \bibfield  {author} {\bibinfo {author} {\bibfnamefont {P.}~\bibnamefont
  {Sala}}, \bibinfo {author} {\bibfnamefont {T.}~\bibnamefont {Rakovszky}},
  \bibinfo {author} {\bibfnamefont {R.}~\bibnamefont {Verresen}}, \bibinfo
  {author} {\bibfnamefont {M.}~\bibnamefont {Knap}}, \ and\ \bibinfo {author}
  {\bibfnamefont {F.}~\bibnamefont {Pollmann}},\ }\href {\doibase
  10.1103/PhysRevX.10.011047} {\bibfield  {journal} {\bibinfo  {journal} {Phys.
  Rev. X}\ }\textbf {\bibinfo {volume} {10}},\ \bibinfo {pages} {011047}
  (\bibinfo {year} {2020})}\BibitemShut {NoStop}%
\bibitem [{\citenamefont {Khemani}\ \emph {et~al.}(2020)\citenamefont
  {Khemani}, \citenamefont {Hermele},\ and\ \citenamefont
  {Nandkishore}}]{Khemani2020}%
  \BibitemOpen
  \bibfield  {author} {\bibinfo {author} {\bibfnamefont {V.}~\bibnamefont
  {Khemani}}, \bibinfo {author} {\bibfnamefont {M.}~\bibnamefont {Hermele}}, \
  and\ \bibinfo {author} {\bibfnamefont {R.}~\bibnamefont {Nandkishore}},\
  }\href {\doibase 10.1103/PhysRevB.101.174204} {\bibfield  {journal} {\bibinfo
   {journal} {Phys. Rev. B}\ }\textbf {\bibinfo {volume} {101}},\ \bibinfo
  {pages} {174204} (\bibinfo {year} {2020})}\BibitemShut {NoStop}%
\bibitem [{\citenamefont {Moudgalya}\ \emph {et~al.}(2021)\citenamefont
  {Moudgalya}, \citenamefont {Prem}, \citenamefont {Nandkishore}, \citenamefont
  {Regnault},\ and\ \citenamefont {Bernevig}}]{Moudgalya2021}%
  \BibitemOpen
  \bibfield  {author} {\bibinfo {author} {\bibfnamefont {S.}~\bibnamefont
  {Moudgalya}}, \bibinfo {author} {\bibfnamefont {A.}~\bibnamefont {Prem}},
  \bibinfo {author} {\bibfnamefont {R.}~\bibnamefont {Nandkishore}}, \bibinfo
  {author} {\bibfnamefont {N.}~\bibnamefont {Regnault}}, \ and\ \bibinfo
  {author} {\bibfnamefont {B.~A.}\ \bibnamefont {Bernevig}},\ }\enquote
  {\bibinfo {title} {{Thermalization and Its Absence within Krylov Subspaces of
  a Constrained Hamiltonian}},}\ in\ \href@noop {} {\emph {\bibinfo {booktitle}
  {Memorial Volume for Shoucheng Zhang}}}\ (\bibinfo {year} {2021})\ Chap.\
  \bibinfo {chapter} {Chapter 7}, pp.\ \bibinfo {pages} {147--209}\BibitemShut
  {NoStop}%
\bibitem [{\citenamefont {De~Tomasi}\ \emph {et~al.}(2019)\citenamefont
  {De~Tomasi}, \citenamefont {Hetterich}, \citenamefont {Sala},\ and\
  \citenamefont {Pollmann}}]{Tomasi2019}%
  \BibitemOpen
  \bibfield  {author} {\bibinfo {author} {\bibfnamefont {G.}~\bibnamefont
  {De~Tomasi}}, \bibinfo {author} {\bibfnamefont {D.}~\bibnamefont
  {Hetterich}}, \bibinfo {author} {\bibfnamefont {P.}~\bibnamefont {Sala}}, \
  and\ \bibinfo {author} {\bibfnamefont {F.}~\bibnamefont {Pollmann}},\ }\href
  {\doibase 10.1103/PhysRevB.100.214313} {\bibfield  {journal} {\bibinfo
  {journal} {Phys. Rev. B}\ }\textbf {\bibinfo {volume} {100}},\ \bibinfo
  {pages} {214313} (\bibinfo {year} {2019})}\BibitemShut {NoStop}%
\bibitem [{\citenamefont {Yang}\ \emph {et~al.}(2020)\citenamefont {Yang},
  \citenamefont {Liu}, \citenamefont {Gorshkov},\ and\ \citenamefont
  {Iadecola}}]{Yang2020}%
  \BibitemOpen
  \bibfield  {author} {\bibinfo {author} {\bibfnamefont {Z.-C.}\ \bibnamefont
  {Yang}}, \bibinfo {author} {\bibfnamefont {F.}~\bibnamefont {Liu}}, \bibinfo
  {author} {\bibfnamefont {A.~V.}\ \bibnamefont {Gorshkov}}, \ and\ \bibinfo
  {author} {\bibfnamefont {T.}~\bibnamefont {Iadecola}},\ }\href {\doibase
  10.1103/PhysRevLett.124.207602} {\bibfield  {journal} {\bibinfo  {journal}
  {Phys. Rev. Lett.}\ }\textbf {\bibinfo {volume} {124}},\ \bibinfo {pages}
  {207602} (\bibinfo {year} {2020})}\BibitemShut {NoStop}%
\bibitem [{\citenamefont {Frey}\ \emph {et~al.}(2022)\citenamefont {Frey},
  \citenamefont {Hackl},\ and\ \citenamefont {Rachel}}]{Frey2022}%
  \BibitemOpen
  \bibfield  {author} {\bibinfo {author} {\bibfnamefont {P.}~\bibnamefont
  {Frey}}, \bibinfo {author} {\bibfnamefont {L.}~\bibnamefont {Hackl}}, \ and\
  \bibinfo {author} {\bibfnamefont {S.}~\bibnamefont {Rachel}},\ }\href
  {\doibase 10.1103/PhysRevB.106.L220301} {\bibfield  {journal} {\bibinfo
  {journal} {Phys. Rev. B}\ }\textbf {\bibinfo {volume} {106}},\ \bibinfo
  {pages} {L220301} (\bibinfo {year} {2022})}\BibitemShut {NoStop}%
\bibitem [{\citenamefont {Ghosh}\ \emph {et~al.}(2023)\citenamefont {Ghosh},
  \citenamefont {Paul},\ and\ \citenamefont {Sengupta}}]{Ghosh2023}%
  \BibitemOpen
  \bibfield  {author} {\bibinfo {author} {\bibfnamefont {S.}~\bibnamefont
  {Ghosh}}, \bibinfo {author} {\bibfnamefont {I.}~\bibnamefont {Paul}}, \ and\
  \bibinfo {author} {\bibfnamefont {K.}~\bibnamefont {Sengupta}},\ }\href
  {\doibase 10.1103/PhysRevLett.130.120401} {\bibfield  {journal} {\bibinfo
  {journal} {Phys. Rev. Lett.}\ }\textbf {\bibinfo {volume} {130}},\ \bibinfo
  {pages} {120401} (\bibinfo {year} {2023})}\BibitemShut {NoStop}%
\bibitem [{\citenamefont {Scherg}\ \emph {et~al.}(2021)\citenamefont {Scherg},
  \citenamefont {Kohlert}, \citenamefont {Sala}, \citenamefont {Pollmann},
  \citenamefont {Hebbe~Madhusudhana}, \citenamefont {Bloch},\ and\
  \citenamefont {Aidelsburger}}]{Kohlert2021}%
  \BibitemOpen
  \bibfield  {author} {\bibinfo {author} {\bibfnamefont {S.}~\bibnamefont
  {Scherg}}, \bibinfo {author} {\bibfnamefont {T.}~\bibnamefont {Kohlert}},
  \bibinfo {author} {\bibfnamefont {P.}~\bibnamefont {Sala}}, \bibinfo {author}
  {\bibfnamefont {F.}~\bibnamefont {Pollmann}}, \bibinfo {author}
  {\bibfnamefont {B.}~\bibnamefont {Hebbe~Madhusudhana}}, \bibinfo {author}
  {\bibfnamefont {I.}~\bibnamefont {Bloch}}, \ and\ \bibinfo {author}
  {\bibfnamefont {M.}~\bibnamefont {Aidelsburger}},\ }\href {\doibase
  10.1038/s41467-021-24726-0} {\bibfield  {journal} {\bibinfo  {journal} {Nat.
  Commun.}\ }\textbf {\bibinfo {volume} {12}},\ \bibinfo {pages} {4490}
  (\bibinfo {year} {2021})}\BibitemShut {NoStop}%
\bibitem [{\citenamefont {Kohlert}\ \emph {et~al.}(2023)\citenamefont
  {Kohlert}, \citenamefont {Scherg}, \citenamefont {Sala}, \citenamefont
  {Pollmann}, \citenamefont {Hebbe~Madhusudhana}, \citenamefont {Bloch},\ and\
  \citenamefont {Aidelsburger}}]{Kohlert2023}%
  \BibitemOpen
  \bibfield  {author} {\bibinfo {author} {\bibfnamefont {T.}~\bibnamefont
  {Kohlert}}, \bibinfo {author} {\bibfnamefont {S.}~\bibnamefont {Scherg}},
  \bibinfo {author} {\bibfnamefont {P.}~\bibnamefont {Sala}}, \bibinfo {author}
  {\bibfnamefont {F.}~\bibnamefont {Pollmann}}, \bibinfo {author}
  {\bibfnamefont {B.}~\bibnamefont {Hebbe~Madhusudhana}}, \bibinfo {author}
  {\bibfnamefont {I.}~\bibnamefont {Bloch}}, \ and\ \bibinfo {author}
  {\bibfnamefont {M.}~\bibnamefont {Aidelsburger}},\ }\href {\doibase
  10.1103/PhysRevLett.130.010201} {\bibfield  {journal} {\bibinfo  {journal}
  {Phys. Rev. Lett.}\ }\textbf {\bibinfo {volume} {130}},\ \bibinfo {pages}
  {010201} (\bibinfo {year} {2023})}\BibitemShut {NoStop}%
\bibitem [{\citenamefont {Morningstar}\ \emph {et~al.}(2020)\citenamefont
  {Morningstar}, \citenamefont {Khemani},\ and\ \citenamefont
  {Huse}}]{Morningstar2020}%
  \BibitemOpen
  \bibfield  {author} {\bibinfo {author} {\bibfnamefont {A.}~\bibnamefont
  {Morningstar}}, \bibinfo {author} {\bibfnamefont {V.}~\bibnamefont
  {Khemani}}, \ and\ \bibinfo {author} {\bibfnamefont {D.~A.}\ \bibnamefont
  {Huse}},\ }\href {\doibase 10.1103/PhysRevB.101.214205} {\bibfield  {journal}
  {\bibinfo  {journal} {Phys. Rev. B}\ }\textbf {\bibinfo {volume} {101}},\
  \bibinfo {pages} {214205} (\bibinfo {year} {2020})}\BibitemShut {NoStop}%
\bibitem [{\citenamefont {Pozderac}\ \emph {et~al.}(2023)\citenamefont
  {Pozderac}, \citenamefont {Speck}, \citenamefont {Feng}, \citenamefont
  {Huse},\ and\ \citenamefont {Skinner}}]{Pozderac2023}%
  \BibitemOpen
  \bibfield  {author} {\bibinfo {author} {\bibfnamefont {C.}~\bibnamefont
  {Pozderac}}, \bibinfo {author} {\bibfnamefont {S.}~\bibnamefont {Speck}},
  \bibinfo {author} {\bibfnamefont {X.}~\bibnamefont {Feng}}, \bibinfo {author}
  {\bibfnamefont {D.~A.}\ \bibnamefont {Huse}}, \ and\ \bibinfo {author}
  {\bibfnamefont {B.}~\bibnamefont {Skinner}},\ }\href {\doibase
  10.1103/PhysRevB.107.045137} {\bibfield  {journal} {\bibinfo  {journal}
  {Phys. Rev. B}\ }\textbf {\bibinfo {volume} {107}},\ \bibinfo {pages}
  {045137} (\bibinfo {year} {2023})}\BibitemShut {NoStop}%
\bibitem [{\citenamefont {Eckardt}(2017)}]{Eckardt2017}%
  \BibitemOpen
  \bibfield  {author} {\bibinfo {author} {\bibfnamefont {A.}~\bibnamefont
  {Eckardt}},\ }\href {\doibase 10.1103/RevModPhys.89.011004} {\bibfield
  {journal} {\bibinfo  {journal} {Rev. Mod. Phys.}\ }\textbf {\bibinfo {volume}
  {89}},\ \bibinfo {pages} {011004} (\bibinfo {year} {2017})}\BibitemShut
  {NoStop}%
\bibitem [{\citenamefont {Greschner}\ \emph {et~al.}(2014)\citenamefont
  {Greschner}, \citenamefont {Santos},\ and\ \citenamefont
  {Poletti}}]{Greschner2014}%
  \BibitemOpen
  \bibfield  {author} {\bibinfo {author} {\bibfnamefont {S.}~\bibnamefont
  {Greschner}}, \bibinfo {author} {\bibfnamefont {L.}~\bibnamefont {Santos}}, \
  and\ \bibinfo {author} {\bibfnamefont {D.}~\bibnamefont {Poletti}},\ }\href
  {\doibase 10.1103/PhysRevLett.113.183002} {\bibfield  {journal} {\bibinfo
  {journal} {Phys. Rev. Lett.}\ }\textbf {\bibinfo {volume} {113}},\ \bibinfo
  {pages} {183002} (\bibinfo {year} {2014})}\BibitemShut {NoStop}%
\bibitem [{\citenamefont {Meinert}\ \emph {et~al.}(2016)\citenamefont
  {Meinert}, \citenamefont {Mark}, \citenamefont {Lauber}, \citenamefont
  {Daley},\ and\ \citenamefont {N\"agerl}}]{Meinert2016}%
  \BibitemOpen
  \bibfield  {author} {\bibinfo {author} {\bibfnamefont {F.}~\bibnamefont
  {Meinert}}, \bibinfo {author} {\bibfnamefont {M.~J.}\ \bibnamefont {Mark}},
  \bibinfo {author} {\bibfnamefont {K.}~\bibnamefont {Lauber}}, \bibinfo
  {author} {\bibfnamefont {A.~J.}\ \bibnamefont {Daley}}, \ and\ \bibinfo
  {author} {\bibfnamefont {H.-C.}\ \bibnamefont {N\"agerl}},\ }\href {\doibase
  10.1103/PhysRevLett.116.205301} {\bibfield  {journal} {\bibinfo  {journal}
  {Phys. Rev. Lett.}\ }\textbf {\bibinfo {volume} {116}},\ \bibinfo {pages}
  {205301} (\bibinfo {year} {2016})}\BibitemShut {NoStop}%
\bibitem [{\citenamefont {Zhao}\ \emph {et~al.}(2019)\citenamefont {Zhao},
  \citenamefont {Knolle},\ and\ \citenamefont {Mintert}}]{Zhao2019}%
  \BibitemOpen
  \bibfield  {author} {\bibinfo {author} {\bibfnamefont {H.}~\bibnamefont
  {Zhao}}, \bibinfo {author} {\bibfnamefont {J.}~\bibnamefont {Knolle}}, \ and\
  \bibinfo {author} {\bibfnamefont {F.}~\bibnamefont {Mintert}},\ }\href
  {\doibase 10.1103/PhysRevA.100.053610} {\bibfield  {journal} {\bibinfo
  {journal} {Phys. Rev. A}\ }\textbf {\bibinfo {volume} {100}},\ \bibinfo
  {pages} {053610} (\bibinfo {year} {2019})}\BibitemShut {NoStop}%
\bibitem [{\citenamefont {Zhao}\ \emph {et~al.}(2020)\citenamefont {Zhao},
  \citenamefont {Vovrosh}, \citenamefont {Mintert},\ and\ \citenamefont
  {Knolle}}]{Zhao2020}%
  \BibitemOpen
  \bibfield  {author} {\bibinfo {author} {\bibfnamefont {H.}~\bibnamefont
  {Zhao}}, \bibinfo {author} {\bibfnamefont {J.}~\bibnamefont {Vovrosh}},
  \bibinfo {author} {\bibfnamefont {F.}~\bibnamefont {Mintert}}, \ and\
  \bibinfo {author} {\bibfnamefont {J.}~\bibnamefont {Knolle}},\ }\href
  {\doibase 10.1103/PhysRevLett.124.160604} {\bibfield  {journal} {\bibinfo
  {journal} {Phys. Rev. Lett.}\ }\textbf {\bibinfo {volume} {124}},\ \bibinfo
  {pages} {160604} (\bibinfo {year} {2020})}\BibitemShut {NoStop}%
\bibitem [{\citenamefont {Liu}\ \emph {et~al.}(2020)\citenamefont {Liu},
  \citenamefont {Ke}, \citenamefont {Zhu},\ and\ \citenamefont
  {Lee}}]{Liu2020}%
  \BibitemOpen
  \bibfield  {author} {\bibinfo {author} {\bibfnamefont {W.}~\bibnamefont
  {Liu}}, \bibinfo {author} {\bibfnamefont {Y.}~\bibnamefont {Ke}}, \bibinfo
  {author} {\bibfnamefont {B.}~\bibnamefont {Zhu}}, \ and\ \bibinfo {author}
  {\bibfnamefont {C.}~\bibnamefont {Lee}},\ }\href {\doibase
  10.1088/1367-2630/abb2b7} {\bibfield  {journal} {\bibinfo  {journal} {New J.
  Phys.}\ }\textbf {\bibinfo {volume} {22}},\ \bibinfo {pages} {093052}
  (\bibinfo {year} {2020})}\BibitemShut {NoStop}%
\bibitem [{\citenamefont {Soori}\ and\ \citenamefont {Sen}(2010)}]{Soori2010}%
  \BibitemOpen
  \bibfield  {author} {\bibinfo {author} {\bibfnamefont {A.}~\bibnamefont
  {Soori}}\ and\ \bibinfo {author} {\bibfnamefont {D.}~\bibnamefont {Sen}},\
  }\href {\doibase 10.1103/PhysRevB.82.115432} {\bibfield  {journal} {\bibinfo
  {journal} {Phys. Rev. B}\ }\textbf {\bibinfo {volume} {82}},\ \bibinfo
  {pages} {115432} (\bibinfo {year} {2010})}\BibitemShut {NoStop}%
\bibitem [{\citenamefont {Sen}\ \emph {et~al.}(2021)\citenamefont {Sen},
  \citenamefont {Sen},\ and\ \citenamefont {Sengupta}}]{Sen2021}%
  \BibitemOpen
  \bibfield  {author} {\bibinfo {author} {\bibfnamefont {A.}~\bibnamefont
  {Sen}}, \bibinfo {author} {\bibfnamefont {D.}~\bibnamefont {Sen}}, \ and\
  \bibinfo {author} {\bibfnamefont {K.}~\bibnamefont {Sengupta}},\ }\href
  {\doibase 10.1088/1361-648X/ac1b61} {\bibfield  {journal} {\bibinfo
  {journal} {J. Phys.: Condens.Matter}\ }\textbf {\bibinfo {volume} {33}},\
  \bibinfo {pages} {443003} (\bibinfo {year} {2021})}\BibitemShut {NoStop}%
\bibitem [{\citenamefont {Page}(1993)}]{Page1993}%
  \BibitemOpen
  \bibfield  {author} {\bibinfo {author} {\bibfnamefont {D.~N.}\ \bibnamefont
  {Page}},\ }\href {\doibase 10.1103/PhysRevLett.71.1291} {\bibfield  {journal}
  {\bibinfo  {journal} {Phys. Rev. Lett.}\ }\textbf {\bibinfo {volume} {71}},\
  \bibinfo {pages} {1291} (\bibinfo {year} {1993})}\BibitemShut {NoStop}%
\bibitem [{\citenamefont {Herviou}\ \emph {et~al.}(2021)\citenamefont
  {Herviou}, \citenamefont {Bardarson},\ and\ \citenamefont
  {Regnault}}]{Herviou2021}%
  \BibitemOpen
  \bibfield  {author} {\bibinfo {author} {\bibfnamefont {L.}~\bibnamefont
  {Herviou}}, \bibinfo {author} {\bibfnamefont {J.~H.}\ \bibnamefont
  {Bardarson}}, \ and\ \bibinfo {author} {\bibfnamefont {N.}~\bibnamefont
  {Regnault}},\ }\href {\doibase 10.1103/PhysRevB.103.134207} {\bibfield
  {journal} {\bibinfo  {journal} {Phys. Rev. B}\ }\textbf {\bibinfo {volume}
  {103}},\ \bibinfo {pages} {134207} (\bibinfo {year} {2021})}\BibitemShut
  {NoStop}%
\bibitem [{\citenamefont {Bravyi}\ \emph {et~al.}(2011)\citenamefont {Bravyi},
  \citenamefont {DiVincenzo},\ and\ \citenamefont {Loss}}]{Bravyi2011}%
  \BibitemOpen
  \bibfield  {author} {\bibinfo {author} {\bibfnamefont {S.}~\bibnamefont
  {Bravyi}}, \bibinfo {author} {\bibfnamefont {D.~P.}\ \bibnamefont
  {DiVincenzo}}, \ and\ \bibinfo {author} {\bibfnamefont {D.}~\bibnamefont
  {Loss}},\ }\href {\doibase https://doi.org/10.1016/j.aop.2011.06.004}
  {\bibfield  {journal} {\bibinfo  {journal} {Ann. Phys.}\ }\textbf {\bibinfo
  {volume} {326}},\ \bibinfo {pages} {2793} (\bibinfo {year}
  {2011})}\BibitemShut {NoStop}%
\bibitem [{\citenamefont {Barredo}\ \emph {et~al.}(2016)\citenamefont
  {Barredo}, \citenamefont {de~L\'{e}s\'{e}leuc}, \citenamefont {Lienhard},
  \citenamefont {Lahaye},\ and\ \citenamefont {Browaeys}}]{Barredo2016}%
  \BibitemOpen
  \bibfield  {author} {\bibinfo {author} {\bibfnamefont {D.}~\bibnamefont
  {Barredo}}, \bibinfo {author} {\bibfnamefont {S.}~\bibnamefont
  {de~L\'{e}s\'{e}leuc}}, \bibinfo {author} {\bibfnamefont {V.}~\bibnamefont
  {Lienhard}}, \bibinfo {author} {\bibfnamefont {T.}~\bibnamefont {Lahaye}}, \
  and\ \bibinfo {author} {\bibfnamefont {A.}~\bibnamefont {Browaeys}},\ }\href
  {\doibase 10.1126/science.aah3778} {\bibfield  {journal} {\bibinfo  {journal}
  {Science}\ }\textbf {\bibinfo {volume} {354}},\ \bibinfo {pages} {1021}
  (\bibinfo {year} {2016})}\BibitemShut {NoStop}%
\bibitem [{\citenamefont {Endres}\ \emph {et~al.}(2016)\citenamefont {Endres},
  \citenamefont {Bernien}, \citenamefont {Keesling}, \citenamefont {Levine},
  \citenamefont {Anschuetz}, \citenamefont {Krajenbrink}, \citenamefont
  {Senko}, \citenamefont {Vuletic}, \citenamefont {Greiner},\ and\
  \citenamefont {Lukin}}]{Endres2016}%
  \BibitemOpen
  \bibfield  {author} {\bibinfo {author} {\bibfnamefont {M.}~\bibnamefont
  {Endres}}, \bibinfo {author} {\bibfnamefont {H.}~\bibnamefont {Bernien}},
  \bibinfo {author} {\bibfnamefont {A.}~\bibnamefont {Keesling}}, \bibinfo
  {author} {\bibfnamefont {H.}~\bibnamefont {Levine}}, \bibinfo {author}
  {\bibfnamefont {E.~R.}\ \bibnamefont {Anschuetz}}, \bibinfo {author}
  {\bibfnamefont {A.}~\bibnamefont {Krajenbrink}}, \bibinfo {author}
  {\bibfnamefont {C.}~\bibnamefont {Senko}}, \bibinfo {author} {\bibfnamefont
  {V.}~\bibnamefont {Vuletic}}, \bibinfo {author} {\bibfnamefont
  {M.}~\bibnamefont {Greiner}}, \ and\ \bibinfo {author} {\bibfnamefont
  {M.~D.}\ \bibnamefont {Lukin}},\ }\href {\doibase 10.1126/science.aah3752}
  {\bibfield  {journal} {\bibinfo  {journal} {Science}\ }\textbf {\bibinfo
  {volume} {354}},\ \bibinfo {pages} {1024} (\bibinfo {year}
  {2016})}\BibitemShut {NoStop}%
\bibitem [{\citenamefont {Browaeys}\ \emph {et~al.}(2016)\citenamefont
  {Browaeys}, \citenamefont {Barredo},\ and\ \citenamefont
  {Lahaye}}]{Browaeys2016}%
  \BibitemOpen
  \bibfield  {author} {\bibinfo {author} {\bibfnamefont {A.}~\bibnamefont
  {Browaeys}}, \bibinfo {author} {\bibfnamefont {D.}~\bibnamefont {Barredo}}, \
  and\ \bibinfo {author} {\bibfnamefont {T.}~\bibnamefont {Lahaye}},\ }\href
  {\doibase 10.1088/0953-4075/49/15/152001} {\bibfield  {journal} {\bibinfo
  {journal} {J. Phys. B: At. Mol. Opt. Phys.}\ }\textbf {\bibinfo {volume}
  {49}},\ \bibinfo {pages} {152001} (\bibinfo {year} {2016})}\BibitemShut
  {NoStop}%
\bibitem [{\citenamefont {Barredo}\ \emph {et~al.}(2018)\citenamefont
  {Barredo}, \citenamefont {Lienhard}, \citenamefont {de~L\'{e}s\'{e}leuc},
  \citenamefont {Lahaye},\ and\ \citenamefont {Browaeys}}]{Barredo2018}%
  \BibitemOpen
  \bibfield  {author} {\bibinfo {author} {\bibfnamefont {D.}~\bibnamefont
  {Barredo}}, \bibinfo {author} {\bibfnamefont {V.}~\bibnamefont {Lienhard}},
  \bibinfo {author} {\bibfnamefont {S.}~\bibnamefont {de~L\'{e}s\'{e}leuc}},
  \bibinfo {author} {\bibfnamefont {T.}~\bibnamefont {Lahaye}}, \ and\ \bibinfo
  {author} {\bibfnamefont {A.}~\bibnamefont {Browaeys}},\ }\href {\doibase
  https://doi.org/10.1038/s41586-018-0450-2} {\bibfield  {journal} {\bibinfo
  {journal} {Nature (London)}\ }\textbf {\bibinfo {volume} {561}},\ \bibinfo
  {pages} {79} (\bibinfo {year} {2018})}\BibitemShut {NoStop}%
\bibitem [{\citenamefont {Browaeys}\ and\ \citenamefont
  {Lahaye}(2020)}]{Browaeys2020}%
  \BibitemOpen
  \bibfield  {author} {\bibinfo {author} {\bibfnamefont {A.}~\bibnamefont
  {Browaeys}}\ and\ \bibinfo {author} {\bibfnamefont {T.}~\bibnamefont
  {Lahaye}},\ }\href {\doibase https://doi.org/10.1038/s41567-019-0733-z}
  {\bibfield  {journal} {\bibinfo  {journal} {Nat. Phys.}\ }\textbf {\bibinfo
  {volume} {16}},\ \bibinfo {pages} {132} (\bibinfo {year} {2020})}\BibitemShut
  {NoStop}%
\bibitem [{\citenamefont {Yang}\ \emph {et~al.}(2022)\citenamefont {Yang},
  \citenamefont {Wang}, \citenamefont {Zhou},\ and\ \citenamefont
  {Liu}}]{Yang2022}%
  \BibitemOpen
  \bibfield  {author} {\bibinfo {author} {\bibfnamefont {T.-H.}\ \bibnamefont
  {Yang}}, \bibinfo {author} {\bibfnamefont {B.-Z.}\ \bibnamefont {Wang}},
  \bibinfo {author} {\bibfnamefont {X.-C.}\ \bibnamefont {Zhou}}, \ and\
  \bibinfo {author} {\bibfnamefont {X.-J.}\ \bibnamefont {Liu}},\ }\href
  {\doibase 10.1103/PhysRevA.106.L021101} {\bibfield  {journal} {\bibinfo
  {journal} {Phys. Rev. A}\ }\textbf {\bibinfo {volume} {106}},\ \bibinfo
  {pages} {L021101} (\bibinfo {year} {2022})}\BibitemShut {NoStop}%
\bibitem [{\citenamefont {Omran}\ \emph {et~al.}(2019)\citenamefont {Omran},
  \citenamefont {Levine}, \citenamefont {Keesling}, \citenamefont {Semeghini},
  \citenamefont {Wang}, \citenamefont {Ebadi}, \citenamefont {Bernien},
  \citenamefont {Zibrov}, \citenamefont {Pichler}, \citenamefont {Choi},
  \citenamefont {Cui}, \citenamefont {Rossignolo}, \citenamefont {Rembold},
  \citenamefont {Montangero}, \citenamefont {Calarco}, \citenamefont {Endres},
  \citenamefont {Greiner}, \citenamefont {Vuleti\'{c}},\ and\ \citenamefont
  {Lukin}}]{Omran2019}%
  \BibitemOpen
  \bibfield  {author} {\bibinfo {author} {\bibfnamefont {A.}~\bibnamefont
  {Omran}}, \bibinfo {author} {\bibfnamefont {H.}~\bibnamefont {Levine}},
  \bibinfo {author} {\bibfnamefont {A.}~\bibnamefont {Keesling}}, \bibinfo
  {author} {\bibfnamefont {G.}~\bibnamefont {Semeghini}}, \bibinfo {author}
  {\bibfnamefont {T.~T.}\ \bibnamefont {Wang}}, \bibinfo {author}
  {\bibfnamefont {S.}~\bibnamefont {Ebadi}}, \bibinfo {author} {\bibfnamefont
  {H.}~\bibnamefont {Bernien}}, \bibinfo {author} {\bibfnamefont {A.~S.}\
  \bibnamefont {Zibrov}}, \bibinfo {author} {\bibfnamefont {H.}~\bibnamefont
  {Pichler}}, \bibinfo {author} {\bibfnamefont {S.}~\bibnamefont {Choi}},
  \bibinfo {author} {\bibfnamefont {J.}~\bibnamefont {Cui}}, \bibinfo {author}
  {\bibfnamefont {M.}~\bibnamefont {Rossignolo}}, \bibinfo {author}
  {\bibfnamefont {P.}~\bibnamefont {Rembold}}, \bibinfo {author} {\bibfnamefont
  {S.}~\bibnamefont {Montangero}}, \bibinfo {author} {\bibfnamefont
  {T.}~\bibnamefont {Calarco}}, \bibinfo {author} {\bibfnamefont
  {M.}~\bibnamefont {Endres}}, \bibinfo {author} {\bibfnamefont
  {M.}~\bibnamefont {Greiner}}, \bibinfo {author} {\bibfnamefont
  {V.}~\bibnamefont {Vuleti\'{c}}}, \ and\ \bibinfo {author} {\bibfnamefont
  {M.~D.}\ \bibnamefont {Lukin}},\ }\href {\doibase DOI:
  10.1126/science.aax9743} {\bibfield  {journal} {\bibinfo  {journal}
  {Science}\ }\textbf {\bibinfo {volume} {365}},\ \bibinfo {pages} {570}
  (\bibinfo {year} {2019})}\BibitemShut {NoStop}%
\end{thebibliography}

%

\end{document}